\documentclass{sig-alternate}

\usepackage{amsmath}
\usepackage{tabularx, booktabs}
\usepackage{url}
\usepackage{eurosym}
\usepackage{multirow}
\usepackage{array}
\usepackage[table]{xcolor}
\usepackage{makeidx}
\usepackage{booktabs}
\usepackage{pifont}
\usepackage{diagbox}
\usepackage{slashbox,booktabs,amsmath}
\usepackage{hhline}

\usepackage[T1]{fontenc}
\usepackage{array}
\usepackage{makecell}
\newcolumntype{x}[1]{>{\centering\arraybackslash}p{#1}}

\usepackage{tikz}

\newcounter{rowcount}
\definecolor{amber}{rgb}{1.0, 0.49, 0.0}
\definecolor{ao(english)}{rgb}{0.0, 0.5, 0.0}

\newcommand{\breakcell}[2][c]{\begin{tabular}[#1]{@{}l@{}}#2\end{tabular}}

\newcommand*\rot{\rotatebox{90}}

\newcommand{\remove}[1]{} 

\newcommand{\replace}[2]{#1} 

\begin{document}

\CopyrightYear{2016}
\setcopyright{acmlicensed}
\conferenceinfo{JCDL '16,}{June 19 - 23, 2016, Newark, NJ, USA}
\isbn{978-1-4503-4229-2/16/06}\acmPrice{\$15.00}
\doi{http://dx.doi.org/10.1145/2910896.2910898}

\title{
Profiling vs. Time vs. Content: What does Matter for Top-k Publication Recommendation based on Twitter Profiles?
}
 
\numberofauthors{2}
\author{
\alignauthor
Chifumi Nishioka\\
       \affaddr{Kiel University, Germany}\\
       \affaddr{ZBW -- Leibniz Information Centre for Economics, Germany}\\
       \email{chni@informatik.uni-kiel.de}
\alignauthor
Ansgar Scherp\\
       \affaddr{ZBW -- Leibniz Information Centre for Economics, Germany}\\
       \affaddr{Kiel University, Germany}\\
       \email{a.scherp@zbw.eu}
}
  
\maketitle
\begin{abstract}
So far it is unclear how different factors of a scientific publication recommender system based on users' tweets have an influence on the recommendation performance. 
We examine three different factors, namely profiling method, temporal decay, and richness of content. 
Regarding profiling, we compare CF-IDF that replaces terms in TF-IDF by semantic concepts, HCF-IDF as novel hierarchical variant of CF-IDF, and topic modeling.
As temporal decay functions, we apply sliding window and exponential decay.
In terms of the richness of content, we compare recommendations using both full-texts and titles of publications and using only titles.
Overall, the three factors make twelve recommendation strategies. 
We have conducted an online experiment with $123$ participants and compared the strategies in a within-group design. 
The best recommendations are achieved by the strategy combining CF-IDF, sliding window, and with full-texts. 
However, the strategies using the novel HCF-IDF profiling method achieve similar results with just using the titles of the publications. 
Therefore, HCF-IDF can make recommendations when only short and sparse data is available. 
\end{abstract} 


\remove{

}

\remove{ }

\section{Introduction}
\label{sec:introduction}
%
The social media platform Twitter is popular among scientists to share and discuss their professional thoughts and interests~\cite{letierce:2010}.
Thus, they are a natural resource for building up a user's professional profile and using it for recommending scientific publications.
Recommending scientific publications based on a user's social media items has several advantages: 
First, users receive recommendations based on their current and ongoing professional interests.
In contrast, systems like Google Scholar\remove{\footnote{\url{http://www.onlinecollege.org/2012/08/17/google-scholars-new-updates-recommended-you/}, last access: 04/01/2016}} and Sugiyama et al.~\cite{sugiyama:2010} recommend scientific publications based on a user's publication record.
It can take up to two years (for conferences) or longer (for journals) until a paper is taken into consideration by the recommender system. 
Second, content-based profiling from a user's social media items mitigates the well-known cold-start problem observed in collaborative filtering systems~\cite{jannach:2010}.
The cold-start problem refers to the initial situation where a recommender system yet does not know anything about a user's interests.
Collaborative filtering systems need to analyze a large amount of user activities in order to provide reasonable recommendations. 
In contrast, content-based recommender systems like our work make recommendations based on similarity scores between a user profile and candidate items. 
Therefore, they can generate recommendations based on a single user profile already. 

There is various research on user profiling from social media items~\cite{chen:2010, orlandi:2012, shen:2013, zhao:2015} and recommending scientific publications~\cite{li:2013,sugiyama:2010,wang:2011}.
However, it is unclear how different profiling methods affect the recommendation performance. 
In addition, the age of social media items as well as scientific publications has an influence on profiling~\cite{shen:2013,orlandi:2012}.
But again, it has not been compared.
Finally, we investigate whether it is possible to make reasonable recommendations when using only the publications' titles, i.\,e., when only short and sparse information about the candidate items is available. 
We have conducted an online experiment to evaluate these three factors of top-$k$ recommendations of scientific publications based on a user's social media profile.
In detail, the factors are:
 
\textbf{(i) Profiling Method:} 
The first factor is the \textit{Profiling Method}, 
where we use Concept Frequency Inverse Document Frequency (CF-IDF)~\cite{goossen:2011} as baseline.
CF-IDF is a modification of TF-IDF where term frequencies are replaced by frequencies of semantic concepts. 
In an experiment with 19 participants, Goossen et al. have shown that CF-IDF outperforms TF-IDF for news article recommendations~\cite{goossen:2011}.
Recently, we have extended the statistical strength of CF-IDF with the semantics provided by a hierarchical knowledge base~\cite{nishioka:2015}. 
The resulting Hierarchical CF-IDF (HCF-IDF) model is capable of revealing semantic concepts that are not explicitly mentioned in texts but still are highly relevant.
This is achieved by applying a spreading activation over a hierarchical knowledge base, which is typically provided as domain-specific taxonomy.
Please note that we also considered using BM25 and TF-IDF as profiling method.
However, our earlier work showed that HCF-IDF performs better for user profiling from social media items~\cite{nishioka:2015}. 
As third method, we apply Latent Dirichlet Allocation (LDA)~\cite{blei:2003, blei:2006}, a state-of-the-art topic modeling method. 
LDA is a generative machine learning approach and thus does not require any prior information such as a knowledge base.
 
\textbf{(ii) Decay Function:}
As the second factor, we investigate two temporal \textit{Decay Function}s.
They are based on the idea that the importance of information declines gradually as time passes. 
We compare sliding window~\cite{shen:2013} and exponential decay~\cite{orlandi:2012,sugiyama:2010}.
Both decay functions have been used in the past for user profiling~\cite{shen:2013,orlandi:2012,sugiyama:2010}.
But so far no comparative study was carried out.

\textbf{(iii) Document Content:}
The third factor defines the richness of \textit{Document Content} used for profiling candidate items (i\,.e., scientific publications).
We compare the use of full-texts and titles of scientific publications for profiling versus profiling only based on titles.

We compared twelve recommendation strategies making use of different combinations of the three factors described above.
For the experiment, we have recruited $n=123$ participants who are posting about their professional interests on Twitter.
For each strategy, the participants have received recommendations of five publications from a large corpus of $|D|=279,381$ scientific publications in the broader field of economics.
We used rankscore~\cite{breese:1998} to measure the recommendation performance. 
We also computed Mean Average Precision (MAP), Precision, Mean Reciprocal Rank (MRR), and normalized Discounted Cumulative Gain (nDCG), which show similar results and documented in the TR~\cite{appendix}. 

The results are very interesting: 
The strategy that employs the profiling method CF-IDF and the decay function Sliding window with both titles and full-texts achieves the overall best recommendation performance.
Although the strategy using CF-IDF shows the highest performance, it has a drawback that it requires full-texts of scientific publications.
Thus, it is remarkable that the strategies with HCF-IDF can achieve comparable results using only titles.
We observe no significant difference between the best performing strategy and strategies with HCF-IDF. 
Thus, we conclude that the use of the spreading activation function over the hierarchical knowledge base enables HCF-IDF to compensate for the sparseness when only titles are available due to e\,.g., legal reasons to hinder the use of full-texts. 
Please note, there is no lack in domain-specific hierarchical knowledge bases such as the one used in the experiment for economics.
In fact, these knowledge bases are freely available for many domains\footnote{\url{http://www.w3.org/2001/sw/wiki/SKOS/Datasets}}.
Furthermore, they are manually crafted by domain experts and thus are of high quality.

\remove{
In addition to the investigated factors above, we have also analyzed the influence of demographic factors (e\,.g., age, gender, academic degree) on the recommendation performance.
The analyses revealed that only the demographic factors gender and academic degree have effects on the recommendation performance. 
Specifically, female participants are significantly more likely to evaluate recommended publications as interesting than male participants. 
In addition, participants whose highest academic degree is Bachelor are significantly more likely to indicate recommended publications as interesting than faculty members.
However, both of the demographic factors are independent from the twelve strategies. 
Thus, they make no difference regarding which strategy performs better.
%
}
In addition, we have applied a correlation analysis between the recommendation performance and the number of tweets a participant has published, the number of concepts extracted from these tweets, the number of concepts extracted per tweet, and the percentage of tweets containing at least one concept respectively.
Our results show no significant correlations in any strategies.
Thus, the methods are robust against the amount of tweets.
  
Subsequently, we review related work in Section~\ref{sec:related-work}. 
Section~\ref{sec:problem-definition} introduces the problem definition.
In Section~\ref{sec:recommender-system}, we describe the three experimental factors used in o4ur recommender system. 
We present the experiment setup and procedure in Section~\ref{sec:evaluation}.
The results are presented in Section~\ref{sec:results} and discussed in Section~\ref{sec:discussion} before we conclude the paper.
\section{Related Work}
\label{sec:related-work}
Recommender systems are categorized into content-based recommender and collaborative filtering~\cite{jannach:2010}. 
Collaborative filtering requires analyzing a large amount of user activities in order to predict items to other users~\cite{zhao:2015}. 
In contrast, we focus on content-based recommender, which suggest items based on similarity scores between a user profile and candidate items. 
A content-based recommender can make recommendations based on data from a single user already. 
Thus it does not suffer from the cold start problem. 
Recommender systems for scientific publications mostly employed user profiles based on publications~\cite{sugiyama:2010,sugiyama:2013} or clicks~\cite{li:2013}.
Instead, we create user profiles based on social media items. 
  
Many works have extracted user interests from social media platforms~\cite{chen:2010, orlandi:2012, shen:2013, zhao:2015}. 
Chen et al.~\cite{chen:2010} studied a recommender system incorporating Twitter, which recommended URLs based on a user's tweets and follower-followee relationships. 
In order to find out the best recommendation strategy, they evaluated twelve strategies from three factors: content sources, topic interest models for users, and social popularity. 
Referring to the factor content sources, Chen et al. showed that profiling based on one's own tweets performed better than based on tweets by one's followees. 
Hence, we build up user profiles from social media items produced by the users themselves. 

In the past years, profiling methods based on semantic concepts (i\,.e., ontology-based profiling) extraction have been developed~\cite{goossen:2011, lu:2012}. 
They extract semantic concepts from texts, using a structured knowledge base, e\,.g., DBpedia\remove{\footnote{\url{http://dbpedia.org/About}, 04/01/2016}}.  
Goossen et al.~\cite{goossen:2011} proposed CF-IDF, as an extention of TF-IDF. 
CF-IDF counts frequencies of a concept instead of a term. 
Their news arcticle recommendation experiment with $19$ participants demonstrated that CF-IDF outperforms TF-IDF. 
Lu et al.~\cite{lu:2012} proposed a recommender system for tweets based on what a user tweeted. 
They constructed user profiles represented as a set of weighted Wikipedia concepts that correspond to Wikipedia articles. 
The experiment demonstrated that concept-based approaches outperform TF-IDF. 
Other works employed a hierarchical structure of a knowledge base for profiling~\cite{kapanipathi:2014, middleton:2001, lu:2012} and demonstrated their effectiveness. 
These approaches can reveal user interests that are not explicitly mentioned in the texts, using a structure of a knowledge base and spreading activation. 
In particular, Middleton et al.~\cite{middleton:2001} constructed user profiles based on a hierarchical knowledge base using spreading activation for a recommender system of scientific publications. 
Their user experiment compared a profiling method using the structure of a hierarchical knowledge base and a method not using the structure. 
The result demonstrated superiority of using the hierarchical knowledge base.  
%
%
Topic modeling such as LDA~\cite{blei:2003} is one of the most popular profiling methods.
It is used in the context of social media~\cite{hong:2010} but particularly suited for document profiling. 

Time-aware user profiles are constructed based on the assumption that the degree of user interests declines as time passes. 
The decline of user interests is modeled by a decay function. 
In the past, the decay functions sliding window~\cite{shen:2013} and exponential decay~\cite{orlandi:2012,sugiyama:2010} have been employed for user profiling. 
However, they have not been compared so far like we do in this work. 
%
%
\section{Problem Definition}
\label{sec:problem-definition}
We address the problem of taking the social media stream as input in order to recommend items such as scientific publications the user might be interested in. 
The problem can be decomposed into three parts:
(1)~First, we need to extract the professional interests that a user exposes through his social media stream and represent the interests in a user profile. 
(2)~Likewise, we profile candidate items (i\,.e., scientific publications) and represent them in a way that they are comparable with the user profile.
(3)~We need a ranking function to compute the top-$k$ items based on similarity scores between the user profile and each candidate item. 
In the following, we formalize the three steps required to create a recommender system based on a user's professional interests extracted from the social media stream.
Symbols used in this paper are summarized in Table~\ref{tb:notation}.

\textbf{(1) User profiling from social media items.}
We consider $I_{u}$ as set of social media items $i$ produced by user $u$. 
A social media item $i \in I_{u}$ has a certain time stamp $t_{i}$.
Subsequently, $P_{u}$, the user profile of the user $u$, is created over a set of concepts $C$ by assigning a specific weight for each concept $c \in C$. 
Generally speaking, a concept $c$ is a key subject in a dedicated field, coming from a given domain-specific knowledge base $C$.  
For instance, ''financial crisis'' is a concept in the field of economics. 
We construct $P_u$ by employing different user profiling functions $\Phi$ and we compare them. 
Formally, user profiles are defined as:
\begin{equation}
\label{eq:user-profile}
P_{u} = \Phi(I_u, C) := \{ (c, w(c , I_u)) \mid \forall c \in C \}  
\end{equation}
Here, $w$ is an arbitrary weighting function that returns a weight of a concept $c$ in a user's social media stream $I_u$.
Thus, it determines how important a concept $c$ is for the user $u$. 
Profiling functions $\Phi$ and weighting functions $w$ are described in Sections~\ref{sec:profiling-method} and \ref{sec:temporal-decay-function}. 
Specifically, we describe weighting functions $w'$ that do not consider temporal decay in Section~\ref{sec:profiling-method} and provide weighting functions $w$ which extend $w'$ with temporal decay in Section~\ref{sec:temporal-decay-function}. 

\textbf{(2) Profiling candidate items.}
We have a set of candidate items $D$. 
A candidate item $d \in D$ has a time stamp $t_{d}$, indicating its published year.  
To determine the similarity scores between a user profile $P_u$ and each candidate item $d \in D$, we need to construct profiles of candidate items in a way that they are comparable with the user profile.  
Formally, we represent a candidate item $d$ as a profile $P_d = \Phi(d, C) := \{(c, w(c , d)) \mid \forall c \in C\}$. 
Since our candidate items are scientific publications, we refer to this process document profiling. 
\begin{table}
\centering
\small
\caption{Symbol Notation}
\begin{tabular}{|c|l|} \hline
    $u$ & a user \\  \hline
		$i$ & a social media item \\  \hline
    $I_u$ & the set of $u$'s social media items \\  \hline
    $c$ & a concept \\  \hline
    $C$ & the set of concepts \\  \hline
    $d$ & a candidate item (scientific publication) \\  \hline
    $D$ & the set of candidate items \\  \hline
		$t_i$, $t_d$ & the time stamp of $i$ and $d$, respectively \\  \hline
		$P_u$ & $u$'s user profile \\  \hline
		$P_d$ & $d$'s document profile \\  \hline
    $\Phi$ & a profiling function \\  \hline
		$w'$ & a weighting function (not considering temporal decay) \\  \hline
		$f$ & a decay function \\  \hline
		$w$ & a weighting function that extends $w'$ with temporal decay\\  \hline
		$\sigma$ & a similarity function \\  \hline
\end{tabular}
\label{tb:notation}
\end{table}
\begin{table}
\small
\centering
\caption{Three factors and their choices for the experiment spanning in total $3 \times 2 \times 2 = 12$ strategies}
	\begin{tabular}{|l|c|c|c|c|c|c|c|c|c|c|c|c|} \hline
    \multicolumn{1}{|c}{\textbf{Factor}} & \multicolumn{12}{|c|}{\textbf{Possible Design Choices}} \\ \hline \hline
    \multicolumn{1}{|l}{\textit{Profiling Method}} & \multicolumn{4}{|c}{CF-IDF} & \multicolumn{4}{|c}{\centering HCF-IDF} & \multicolumn{4}{|c|}{LDA}  \\  \hline
    \multicolumn{1}{|l}{\textit{Decay Function}} & \multicolumn{6}{|c}{\centering Sliding window} & \multicolumn{6}{|c|}{\centering Exponential decay}  \\  \hline
    \multicolumn{1}{|l}{\textit{Document Content}} & \multicolumn{6}{|c}{\centering All (title + full-text)} & \multicolumn{6}{|c|}{\centering Title} \\ \hline
  \end{tabular}
  \label{tb:factor}
\end{table}

\textbf{(3) Ranking candidate items.}
We rank candidate items based on similarity scores between the user profile $P_{u}$ and a document profile $P_{d}$. 
A similarity function $\sigma$ takes as input a user profile $P_{u}$ and document profile $P_{d}$. 
It is defined as $\sigma(P_{u}, P_d) \rightarrow [0, 1]$. 
The similarity function is applied to all candidate items $d \in D$.
Finally, the top-$k$ most relevant items (i\,.e., documents whose similarity scores with $P_{u}$ are ranked in the top-$k$) are recommended to the user $u$. 
The similarity functions $\sigma$ are described in Section~\ref{sec:similarity-functions}. 

\section{Experimental Factors}
\label{sec:recommender-system}
According to the three factors (i)-(iii) stated in the introduction, we form the design space of our experiment. 
We illustrate the design space in Table~\ref{tb:factor}, where each cell is a possible design choice we can make in one of the three factors. 
Subsequently, we detail the factor \textit{Profiling Method} in Section~\ref{sec:profiling-method} and the factor \textit{Decay Function} in Section~\ref{sec:temporal-decay-function}. 
Further, we describe similarity functions $\sigma$ in Section~\ref{sec:similarity-functions}. 
The factor \textit{Document Content} investigates whether full-texts of scientific publications enhance the recommendation performance compared to using only titles.

\subsection{Profiling Method}
\label{sec:profiling-method}
We investigate three methods for user profiling and document profiling. 
For each method, we define a weighting function $w'$ that gives a certain weight to each concept $c$. 
The final weighting function $w$ taking temporal decay into account is described in Section~\ref{sec:temporal-decay-function}. 

\textbf{CF-IDF:}
Compared to the traditional TF-IDF, CF-IDF (Concept Frequency Inverse Document Frequency) counts frequencies of a semantic concept instead of term frequencies~\cite{goossen:2011}. 
Semantic concepts or short concepts are stored in an external knowledge base.
Each concept has a unique resource identifier (URI) and one or more labels describing the concept~\footnote{\url{https://www.w3.org/DesignIssues/LinkedData.html}}. 
The concept's labels are treated as synonyms.
As an example, the concept ``clothing industry'' has the URI \texttt{http://zbw.eu/stw/version/latest/descriptor/13128-2} and is defined in the thesaurus STW, a domain-specific knowledge base for economics (described in Section~\ref{sec:dataset}). 
The concept has not only the label ``clothing industry'' but also the synonymous labels ``garment industry'' and ``apparel industry''. 
We count the label frequency, i\,.e., the number of times the label appears, in the social media items and candidate items. 
Subsequently, we calculate the concept frequency, i\,.e., the number of times the concept appears, by summing up the frequencies of the labels referring to the concept. 
For instance, if the labels ``clothing industry" and ``garment industry" appear twice and once in a text, the total frequency of the concept referring to ``clothing industry" is three. 

For the social media items $I_u$ of the user $u$, CF-IDF is computed along with Equation~\ref{eq:cf-idf-item}.
\begin{equation}
\label{eq:cf-idf-item}
w'_{cf\text{-}idf} (c , i)=cf(c , i) \cdot \log \frac{|I_{u}|+|I_{r}|}{|\{i \in I_{u} \cup I_{r} : c \in i\}|},
\end{equation}
where $cf(c , i) = \frac{\text{the number of times concept}~c~\text{appears in}~i}{\text{the number of times all concepts appear in}~i}$.
The denominator $|\{i \in I_{u} \cup I_{r} : c \in i\}|$ counts the number of social media items that contain a concept $c$.
$I_r$ is a set of random social media items.

We employ a set of random social media items $I_{r}$, because it allows to better distinguish the relevant concepts in the user's social media items $I_{u}$, as Chen et al.~\cite{chen:2010} and Lu et al.~\cite{lu:2012} did for TF-IDF. 
For instance, assuming there are two social media items from a user $u$ and both include the concept ``currency competition''. 
Although ``currency competition'' should have a high weight in the user profile, in this case IDF and a final CF-IDF score would be $0$ because ``currency competition'' is common in a user $u$'s social media items. 
The random social media items are sampled from public microblog postings.
In our case, they are obtained from the public Twitter stream using the Twitter API.

We have conducted a simple pre-experiment to empirically determine the optimal amount of random tweets to be used in the profiling method in the context of our experiment of recommending economics publications.
Given this pre-experiment, we set the size of random social media items to five times of $|I_u|$.
In more detail, we applied different sizes of $I_r$, starting from $0$ to $1000$ random tweets. 
For $26$ Twitter accounts, we computed the IDF scores for user profile over $I_u \cup I_r$ and compared it using cosine similarity with the user profile computed only over $I_u$.
The Twitter accounts were taken from a list of famous economists\footnote{\url{http://www.huffingtonpost.com/2012/11/13/economists-twitter_n_2122781.html}\remove{, last access: 04/01/2016}} that are frequently tweeting. 
We ensured that the set of random tweets $I_r$ is disjoint do the user's tweets, i.\,e.. $I_r \cap I_u = \emptyset$.
Particularly, we looked into the changes of the cosine similarity while adding more random tweets.
We observed the changes in the IDF scores became stable after about a factor of five w.r.t. to $|I_u|$. 
The changes indicate the influence of the IDF scores to user profile.
%
Using this technique is effective as the IDF score ensures that too generic concepts do not get too high weights in the user profiling. 
Those generic concepts are at the upper levels of the hierarchy of the domain-specific knowledge base. 
In our case those concepts are like ``product'' and ``economics''. 
Please note that the factor may depend on the domain of economics considered in this paper and that a different factor may be chosen for other domains.


Regarding document profiling, CF-IDF is computed as defined in Equation \ref{eq:cf-idf-doc}.
The computation is basically identical with the one for user profiling shown in Equation~\ref{eq:cf-idf-item}. 
The difference is that CF is computed over single documents and IDF is computed over the document collection. 
\begin{equation}
\label{eq:cf-idf-doc}
w'_{cf\text{-}idf}(c , d)=cf(c , d) \cdot \log \frac{|D|}{|\{d \in D\ : c \in d\}|}
\end{equation}

\begin{figure}
\centering
\includegraphics[width=0.8\columnwidth]{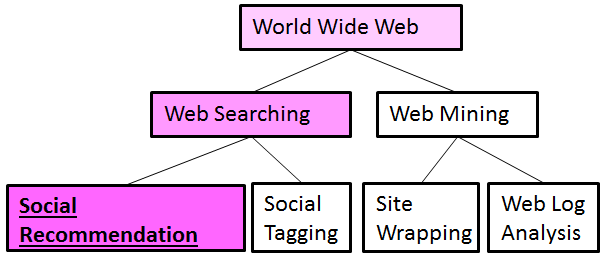}
\caption{An example of HCF-IDF}
\label{fig:spreading-activation}
\end{figure}

\textbf{HCF-IDF:} 
The novel profiling method HCF-IDF (Hierarchical CF-IDF)~\cite{nishioka:2015} extends CF-IDF by using a hierarchical knowledge base, where the concepts are hierarchically organized in a taxonomy. 
HCF-IDF can reveal concepts that are indirectly mentioned in texts by applying a spreading activation over the hierarchical knowledge base. 
Figure~\ref{fig:spreading-activation} shows an example where a user's profile includes the concept ``social recommendation''.
Due to the hierarchical structure of the knowledge base, also the concepts ``web searching'' and ``world wide web'' are activated and obtain non-zero weights even if they are not mentioned. 
Different from the profiling methods using spreading activation~\cite{kapanipathi:2014, middleton:2001}, HCF-IDF avoids to provide too high weights to generic concepts like ``economy'', as it employs IDF. 
Specifically, HCF-IDF combines the statistical strength of CF-IDF with semantics provided by the hierarchical knowledge base. 
We compute HCF-IDF along with Equation~\ref{eq:hcfidf}. 
\begin{equation}
\label{eq:hcfidf}
w'_{hcf\text{-}idf}(c, i)=BL(c, i) \cdot log \frac{|I_u|+|I_r|}{|\{i \in I_u \cup I_r: c \in i\}|}
\end{equation}
$BL(c,i)$ denotes the spreading activation function BellLog from Kapanipathi et al.~\cite{kapanipathi:2014}.
It returns a weight of a concept $c$ in a social media item $i$ and is defined below: 
\begin{equation}
\label{eq:belllog}
BL(c, i) = cf(c, i) + FL(i) \cdot \sum_{c_j \in C_l(c)}{BL(c_j, i)}, 
\end{equation}
where $FL(c) = \frac{1}{\log_{10}(nodes(h(c) + 1))}$. 
$h(c)$ returns the level where a concept $c$ is located in the knowledge base and $nodes$ provides the number of concepts at a given level in a knowledge base. 
For example, in Figure~\ref{fig:spreading-activation} $h(\text{``web searching'})$ returns $2$ and $nodes(h(\text{``web searching'}) + 1)$ returns $4$.
$C_l(c)$ returns the set of concepts located in one level lower than the concept $c$. 
In Figure~\ref{fig:spreading-activation} the function $C_l(\text{``world wide web''})$ returns ``web searching'' and ``web mining''.

For scientific publications, weights are computed as defined in Equation~\ref{eq:hcf-idf-doc}. 
The computation is basically identical with the one for user profiling as shown in Equation~\ref{eq:hcfidf}.
The difference is that $BL$ is applied over single documents and IDF is computed over the document collection. 
\begin{equation}
\label{eq:hcf-idf-doc}
w'_{hcf\text{-}idf}(c, d)=BL(c, d) \cdot log \frac{|D|}{|d \in D: c \in d|}
\end{equation}

\textbf{LDA:}
As third profiling method, we use LDA~\cite{blei:2003, blei:2006}, an unsupervised topic modeling method. 
LDA identifies latent topics in a document collection, where each document is represented as a probability distribution over topics, while each topic is again represented as a probability distribution over a number of words. 
Please note that for user profiling, we treat the set of social media items $I_{u}$ published by a user $u$ as one \emph{single} social media document in this profiling method. 
It is known that topic models that treat a user's microblog postings as one combined social media document outperform topic models computed over single postings of a user for recommendation tasks~\cite{hong:2010}.
We first create a topic model for the entire document collection $D$ (using the parameters and tools described in detail in Section~\ref{sec:dataset}). 
Subsequently, we run LDA with the given topic model for the document collection $D$ and infer a probability distribution over topics for the user's social media document $I_{u}$. 

Again, we use the same notation of concepts $c$ as introduced above: 
Each topic generated by LDA is treated as a concept $c \in C$.
The weight of a concept $c$ is defined by $w'_{lda}(c , I_{u})=p(c \mid I_{u})$ for user profiles and $w'_{lda}(c , d)=p(c \mid d)$ for document profiles, where $p(c \mid d)$ and $p(c \mid I_u)$ denote the probability of the concept (i\,.e., topic) $c$ in the social items $I_{u}$ and document $d$, respectively. 
\subsection{Decay Function}
\label{sec:temporal-decay-function}
We compare two decay functions $f$, namely sliding window and exponential decay. 
In the past, both functions have been used in recommender systems~\cite{shen:2013, orlandi:2012, sugiyama:2010}. 
However, so far they have not been empirically compared.
The profiling functions $w'$ described in the previous section are combined with a decay function $f$ in order to obtain a final weight $w$.
The final weights are computed by Equation~\ref{eq:final-score-i} for the set of social media items and Equation~\ref{eq:final-score-d} for the candidate items.
\begin{equation}
\label{eq:final-score-i}
w(c , I_u)= \sum_{c \in i : i \in I_u} f(t_{i}) \cdot w'(c , i)
\end{equation}
\begin{equation}
\label{eq:final-score-d}
w(c , d)= f(t_{d}) \cdot w'(c , d)
\end{equation} 
Please note that when employing LDA, the decay functions can only be applied on the candidate items, because we treat the user's social media items as one single document.

\textbf{Sliding Window:} 
There are two kinds of sliding window functions, whose window size is defined by (a) the number of items~\cite{khribi:2008} and (b) the period of time~\cite{soltysiak:1998}. 
The approach (a) is employed to identify relatively short-term features (e\,.g., user interests from web browsing histories)~\cite{khribi:2008}, while the approach (b) is used to identify long-term features~\cite{soltysiak:1998}. 
We aim at extracting a user's professional interests, which are rather long-term.
Thus, we take the approach (b) and use only social media items and documents that are younger than a given threshold point in time $thresh$. 
The sliding window function can be represented as Equation \ref{eq:sliding}.
\begin{equation}
\label{eq:sliding}
f_{sw}(t)=\begin{cases}
\begin{array}{ll}
1 & for~t \geq thresh \\
0 & for~t < thresh \\
\end{array}
\end{cases}
\end{equation}
For user profiles, we set the threshold based on the work by Orlandi et al.~\cite{orlandi:2012}. 
They found out that the half life time is $thresh_{social} = 250~days$. 
%
For document profiles, Sangam et al.~\cite{sangam:2013} observed that the half-life time of the scientific publications in the field of social science is $9.04~years$. 
In our experiment, we use a dataset of scientific publications in economics (see Section~\ref{sec:dataset}), which has a large overlap with social science. 
Thus, we set $thresh_{doc} = 9.04~years$~\cite{sangam:2013} and remove scientific publications published more than $9.04~years$ ago from the candidate items. 

\textbf{Exponential Decay:} 
The exponential decay function is defined as shown in Equation~\ref{eq:exponential}. 
\begin{equation}
\label{eq:exponential}
f_{exp}(t)=e^{-(t_{current} - t) / \tau},
\end{equation}
where $t_{current}$ denotes the current time and $\tau$ is a positive number presenting mean-life~\cite{orlandi:2012}.
For user profiles, we set $\tau=360~days$ based on Orlandi~\cite{orlandi:2012}. 
Since Sangam et al.~\cite{sangam:2013} found out that the mean-life of scientific publications in social sciences is $13.05~years$, we set $\tau=13.05~years$ for document profiles. 
\subsection{Similarity Functions}
\label{sec:similarity-functions}
We calculate the similarity scores between a user profile $P_{u}$ and each document profile $P_{d}$. 
We cast a user profile $P_u$ and document profiles $P_d$ to a user profile vector $\vec{p}_{u}$ and document profile vectors $\vec{p}_{d}$, respectively. 
Each element in the vectors corresponds to a weight of a concept $c$. 

\textbf{Temporal Cosine Similarity:}
We employ the temporal cosine similarity function described in Equation~\ref{eq:temporal-cosine-similarity} for the profiling methods CF-IDF and HCF-IDF. 
\begin{eqnarray}
\label{eq:temporal-cosine-similarity}
\sigma_{tcossim}({P}_{u}, {P}_{d}) = f(t_{d}) \cdot \frac{\vec{p_{u}} \cdot \vec{p}_{d}}{{||\vec{p_{u}}||}\cdot{||\vec{p}_{d}||}}, 
\end{eqnarray}
It extends the cosine similarity by the function $f(t_d)$, which results in higher similarity score to newer documents. 
$f(t_d)$ is a decay function from Equation~\ref{eq:sliding} or Equation~\ref{eq:exponential}. 
$t_d$ is time stamp of a scientific publication $d$. i\,.e., the year at which $d$ was published. 
%
%

\textbf{Dot Product:}
For LDA, we employ the dot product computed as $\sigma_{dp}({p}_{u},{p}_{d})=\vec{p}_{u} \cdot \vec{p}_{d}$. 
Since LDA represents documents as probability distribution, it is more reasonable to use Kullback-Leibler divergence (KL divergence). 
However, the dot product outperforms cosine similarity and Kullback-Leibler divergence (KL divergence) when representing documents using LDA~\cite{hazen:2010}. 
\begin{figure*}
\centering
\includegraphics[width=0.8\textwidth]{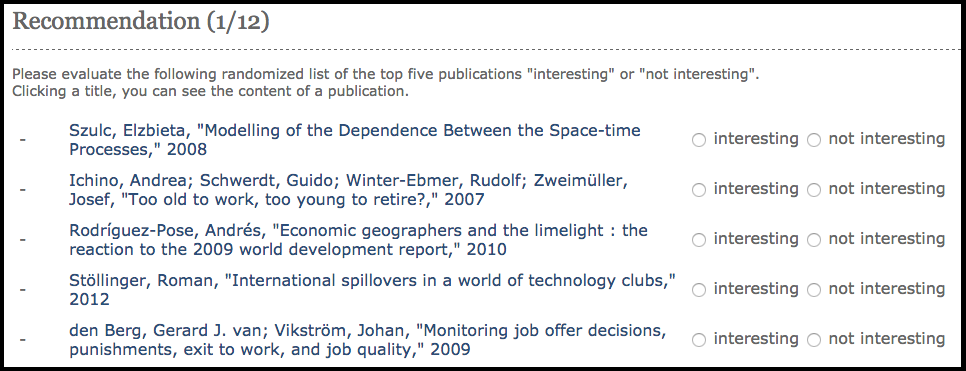}
\caption{Screenshot of our experiment web page showing a randomized list of top-5 recommendations for the first of twelve strategies (which again are randomly ordered). For each recommendation the participants could assess the bibliographic record as well as click on the full-text document. The participants rated each recommended publication as ``interesting'' or ``not interesting''' based on their research interests.}
\label{fig:screenshot}
\end{figure*}
\section{Evaluation}
\label{sec:evaluation}
We conducted an online experiment with $n=123$ participants in order to identify the best strategy for a recommender system along the factors described in Section~\ref{sec:recommender-system}. 
As social media platform, we choose Twitter as it is widely used in scientific communities~\cite{letierce:2010}. 
We design our experiment following the experiment setup and procedure of Chen et al.~\cite{chen:2010}:
Each participant obtains top-$5$ recommendations for each of the twelve strategies formed from the three factors.  
The recommendation performance of each strategy is measured by the rankscore~\cite{breese:1998}. 
%
Below, we describe the details of our experiment procedure and participants. 
Subsequently, we explain the dataset and the knowledge base used in the experiment.
Finally, we introduce our evaluation metric.
\remove{
\subsection{Pilot Studies}
\label{sec:pilot-studies}
We conducted pilot studies with five participants in order to fine-tune the usability of the experiment and to obtain qualitative feedback about the usefulness of the recommendations.
In addition, we quantitatively investigated the potential duplicates of recommended documents by the different strategies as described in Table~\ref{tb:factor}.
Duplicate recommendations may annoy participants throughout the experiment, since they have to evaluate the same items several times under strategies. 
In order to estimate the quantity of duplicate recommendations made, we computed the top-5 recommendation lists of each twelve strategies for 26 famous economists' Twitter accounts\footnote{Famous economists on Twitter: \url{http://www.huffingtonpost.com/2012/11/13/economists-twitter_n_2122781.html}\remove{, last access: 04/01/2016}}. 
In result, on average $76.32$\% (SD: $5.29$) of listed publications are unique.
When conducting a pairwise comparison of the different recommendation factors, we obtain even higher values of uniqueness such as recommendations between no-decay and Exponential decay differ by $87.01$\% (SD: $2.64$).
The recommended document sets between CF-IDF and LDA and HCF-IDF and LDA are almost totally exclusive each other (the uniqueness is 100\% (SD: $0.00$) and 99.92\% (SD: $0.38$), respectively). Regarding CF-IDF and HCF-IDF, they differ by $91.86$\% (SD: $3.56$). In terms of titles and All, the difference between them is $88.98$\% (SD: 1.83). 
}
\subsection{Procedure}
\label{sec:procedure}
The participants are invited to a web application implementing the twelve recommendation strategies. 
First, participants input their public Twitter handles and e-mail address. 
Then, the participants' tweets are retrieved from the Twitter API. 
Subsequently, user profiles are created from the tweets using each of the three profiling methods and two decay functions.
Based on the user profiles, personalized top-$k$ recommendations of scientific publications are generated for each of the twelve strategies.
We set the number of recommendations per strategy $k = 5$ along with Chen et al.~\cite{chen:2010}. 
After computing the recommendations, the participants receive an e-mail invitation to assess the recommendations.
The participant go through all of the twelve strategies like as Chen et al.~\cite{chen:2010}. 
Thus, we apply a repeated measures design.
Each participant obtains $12 \cdot 5 = 60$ recommendations in total throughout the experiment. 
%

Prior to starting the experiment, participants are informed about the task of the experiment, i\,.e., rating the recommended publications based on relevance to their research interests, and confirmed consent. 
On each of the subsequent pages, the participants see a list of five recommendations produced by one of the twelve strategies.
An example screenshot of the evaluation page is shown in Figure~\ref{fig:screenshot}.

For each recommended scientific publication, the participants see its bibliographic information, i\,.e., authors, title, and year of publication. 
In addition, participants can look into the original PDF files by clicking on a link attached to the bibliographic record. 
In order to avoid bias, the participants go through the twelve strategies in random order. 
For each strategy, the participants receive one list of five recommendations.
The five recommendations in the lists are again shown in random order to the participants to avoid the well-known ranking bias.
Typically, participants assume that top-ranked recommendations are essentially more relevant~\cite{bostandjiev:2012,chen:2010}.  
Thus, again prior to starting the experiment we have explicitly informed the participants that we have randomized the order of the items in the top-5 lists.
However, the actual ranks of the recommendations as well as their positions where the recommended items appeared on the participants' screen are stored in the database for later analyses.
Participants evaluate each recommendation as ``interesting" or ``not interesting" by clicking on radio buttons next to the publication records like Chen et al.~\cite{chen:2010}. 
Please note, the participants had to evaluate all recommended items. 

At the end of the experiment, we collect the demographic information of each participant, including gender, age, highest academic degree, major, years of profession, and current employment status (academia/industry).
Finally participants could state free comments regarding the experiment. 
\subsection{Participants}
\label{sec:participants}
We recruited $n=123$ participants through mailing lists, tweets, and word-of-mouth on the Internet. 
Initially $160$ participants registered their Twitter handles and email address for our experiment. 
Among them, $134$ participants started the experiment after receiving the e-mail invitation.
From these $134$ participants, only eleven dropped in the course of assessing the recommendations in the twelve strategies.
%
Thus, finally we obtain evaluations for all strategies from $n=123$ participants.
From these, $27$ participants are female.
The average age of the participants is $32.83$~years (SD:~$7.34$). 
Regarding the highest academic degree, we have acquired $21$ with a Bachelor, $58$ have a Master, $32$ a PhD, and $12$ are lecturers/professors. 
While $83$ participants work in academia, $40$ work in industry. 
Tweets of the participants were retrieved via Twitter API. 
We only collected tweets in English as the scientific publications are also in English. 
The participants published on average $1096.82$ English tweets (SD: $1048.46$). 
The maximum and minimum numbers of tweets are $3192$ and $2$, respectively. 
Twitter users who have not produced any tweets in the last $250~days$ could not register and participate in the experiment, since we use a $250~days$ threshold for the decay function Sliding window (see Section~\ref{sec:temporal-decay-function}). 
Five Twitter users could not participate in the experiment for this reason. 

The participants spent on average $517.54$ seconds to complete the assessment of the $5 \times 12 = 60$ recommendations (SD:~$376.72$). 
This does not include the time spent to register for the experiment, read the instructions, and filling out the final questionnaire. 
As incentive, each participant received the information about his most similar economist among 26~famous economists\footnote{\url{http://www.huffingtonpost.com/2012/11/13/economists-twitter_n_2122781.html}\remove{, last access: 04/01/2016}} 
and the top-$5$ dominant semantic concepts in their tweets after the experiment. 
In addition, the participants could opt-in to a raffle for one of two Amazon vouchers worth of 50~\euro. 
\subsection{Dataset Preparation}
\label{sec:dataset}
We use a large-scale dataset of scientific publications in the field of economics as candidate items and a high-quality taxonomy as a knowledge base for profiling methods. 

\textbf{Dataset of Scientific Publications.}
We collaborate with the providers of EconBiz\footnote{\url{http://www.econbiz.de/}\remove{, last access: 04/01/2016}}, a portal for scientific publications in economics managed by ZBW, the German National Library of Economics. 
From this portal, we obtained 1 million URLs of open access publications and extracted full-texts and metadata (i\,.e., authors, title, year of publication) of $413,098$ scientific publications. 
Finally, we determined the document language\footnote{\url{https://code.google.com/p/language-detection/}\remove{, last access: 04/01/2016}} and got $279,381$ scientific publications in English, which were used in this experiment. 
%

\textbf{Knowledge Base in Economics.} 
The ZBW also maintains and further develops the hierarchical knowledge base STW\footnote{\url{http://zbw.eu/stw/version/8.12/about.en.html}\remove{, last access: 04/01/2016}}, a thesaurus specialized for the field of economics. 
The STW is freely available and is of high quality due to its manual maintenance by domain experts.
The knowledge base is poly-hierarchically organized with six levels.
It contains $6,335$ semantic concepts and $11,679$ labels. 
The hierarchically organized concepts are connected with each other via $14,875$ edges. 
In order to extract as many labels as possible, we enhanced the original \replace{STW}{\textit{Anonymized Thesaurus}} with DBpedia redirects\footnote{\url{http://oldwiki.dbpedia.org/Downloads39#redirects}\remove{, last access: 04/01/2016}}. 
From DBpedia redirects we can retrieve the synonymous labels for a concept.
STW contains $2,692$ concepts that have both a DBpedia mapping and one or more DBpedia redirects.
As an example, for the concept ``Telecommunications industry'' in the thesaurus, we obtain the DBpedia redirects ``Telecommunications operator'' and ``Telephone companies'' and use them as synonymous labels referring to the concept ``Telecommunications industry''.
Finally, our extended \replace{STW}{\textit{Anonymized Thesaurus}} contains $6,335$ concepts and $37,733$ labels. 
This extended STW is used for the profiling methods CF-IDF and HCF-IDF. 
For CF-IDF, we ignore the edges between concepts. 

\textbf{Processing of the tweets and publications.} 
For the profiling methods CF-IDF and HCF-IDF, we extract semantic concepts from the participants' tweets and the scientific publications by matching the texts with the labels from the extended \replace{STW}{\textit{Anonymized Thesaurus}} (i.\,e., a gazetteer-based approach). 
%
Before processing, we lemmatize both the tweets and the scientific publications using Stanford Core NLP\footnote{\url{http://nlp.stanford.edu/software/corenlp.shtml}\remove{, last access: 04/01/2016}} and remove stop words. 
Regarding the tweets, some of them contain hashtags indicating topics (e\,.g., \#election) and user mentions (e\,.g., @UNICEF). 
We remove only the symbols \# and @ from the tweets as Feng et al.~\cite{feng:2014} observed that the combination of the tweets' textual content with the hashtags and user mentions made the highest performance for tag recommendation.

This process extracts only the users' professional interests from tweets and helps to avoid noise (i\,.e., topics not relevant to professional interests in economics). 
A participant has published on average $1096.82$ tweets (SD: $1048.46$). 
On average $1,214.93$ concepts (SD: $1181.43$) are contained in a participant's tweets and $1.07$ concepts (SD: $0.31$) are contained per tweet. 
Regarding CF-IDF and HCF-IDF, we calculate the ratio of the number of tweets containing at least one concept and the total number of tweets the user has published. 
This indicates the percentage of tweets that have contributed to creating the user profile.
On average, $62.24$\% of the tweets (SD: $13.55$) that a participant has published contain at least one concept in economics. 
These tweets are assumed to be relevant to the professional interests. 

\textbf{LDA.} 
For constructing profiles by LDA, we use JGibbLDA\footnote{\url{http://jgibblda.sourceforge.net/}\remove{, last access: 04/01/2016}}.
%
We first run LDA to generate the topic model based on the given document set $D$. 
Following Blei et al.~\cite{blei:2006}, we lemmatize the scientific publications using Stanford NLP Core.
Subsequently, we remove stop words and words that appear in fewer than $25$ scientific publications.
We optimized the number of topics $K$ regarding the maximum mean log likelihood of words given topics as suggested by Griffiths et al.~\cite{griffiths:2004}.
We experimented with $K = 20$, $50$, $100$, $200$, $500$, $1000$, and $5000$ and obtained the highest log likelihood for $K = 100$.
All topic models were computed over $500$ iterations.
Regarding the further parameters for LDA, we set $\alpha = 0.5$ and $\beta = 0.1$ as suggested by Griffiths et al.~\cite{griffiths:2004}.
To infer a topic distribution over a user's tweets, we run LDA again using the topic model for the document set $D$ with $200$ iterations. 
Prior to this, we prepare the tweets of a user $u$ in a single social media document as described in Section~\ref{sec:profiling-method}.
%
\subsection{Evaluation Metric} 
In order to assess the recommendation performance, we compute the rankscore~\cite{breese:1998} as used by Bostandjiev et al.~\cite{bostandjiev:2012} and introduced by Jannach et al~\cite{jannach:2010}. 
Rankscore posits that each successive item in a list is less likely to be viewed by users with an exponential decay, as defined in Equation~\ref{eq:rank-score}. 
\begin{equation}
\label{eq:rank-score}
rankscore' = \sum_{d \in hits}  \frac{1}{2^{\frac{rank_{d} - 1}{\theta - 1}}}
\end{equation}
$\theta$ denotes a viewing halflife parameter controlling the speed of the exponential decay. 
As suggested by Breese et al.~\cite{breese:1998}, we set $\theta=5$. 
$hits$ refers to the set of documents $d$ evaluated as ``interesting'' and $rank_{d}$ denotes the rank of a recommended item $d$ in a list.
Please note $rank_{d}$ denotes the actual rank stored in the database different from the position where a item $d$ appears in the list (cf.~Section~\ref{sec:procedure}).
The normalized rankscore is computed by $rankscore = rankscore' \\ / rankscore_{max}$, where the maximum rankscore $rankscore_{max} = \sum_{j = 1}^{k}\frac{1}{2^{\frac{j - 1}{\theta - 1}}}$. 
Here, $k$ is the number of the recommended items. 
We set $k = 5$. 
%
%
We also computed Mean Average Precision (MAP), Precision@5, Mean Reciprocal Rank (MRR), and normalized Discounted Cumulative Gain (nDCG). 
Overall, the results are similar to the rankscore and thus omitted for reasons of brevity.
The interested reader may refer to the details in the appendix~\cite{appendix}. 
\section{Results}
\label{sec:results}
In this section, we document the results of the experiment\footnote{The anonymized experimental data is available from: \url{http://dx.doi.org/10.7802/1224}} and conduct the statistical analyses. 
We set a significance level of $\alpha = 5\%$ for all statistical tests (please do not confuse with $\alpha$ for LDA in Section~\ref{sec:dataset}).
\subsection{Quantitative Analyses}
\label{sec:quantitative-analyses}
We first report the best performing strategy among the twelve strategies. 
Subsequently, we analyze the influence by the experimental factors followed by investigating the correlations between the recommendation performance and the numbers of tweets written by a user. 
Finally, we analyze the performance related to the number of times the participants clicked on the full-text of a publication. 

\textbf{Best performing strategy.}
Table~\ref{tb:single-all} documents the average rankscores of the twelve strategies sorted in decreasing order. 
Overall, the best performing strategy is the strategy CF-IDF $\times$ Sliding window $\times$ All.
We apply a one-way repeated-measure ANOVA in order to identify if there are significant differences between the strategies. 
For using ANOVA, we first need to verify whether the variances of the rankscores of the twelve strategies are equal. 
This is done by using Mauchly's test, which reveals a violation of sphericity in the strategies ($\chi^2(65) = 435.90$, $p = .00$).
It may lead to positively biased F-statistics and increases the risk of false positives. 
To reduce this risk, we apply a Greenhouse-Geisser correction of $\epsilon = .61$ and run the one-way repeated-measure ANOVA.
It reveals a significant difference in the rankscores of the strategies ($F(6.60, 805.33) = 21.98$, $p = .00$). 
To assess the pair-wise significant differences between the twelve strategies, a post-hoc analysis is conducted.
We have applied Shaffer's modified sequentially rejective Bonferroni procedure (Shaffer's MSRB procedure)~\cite{shaffer:1986} that takes into account the number of different experiment conditions, i.\,e., the number of recommendation strategies.
The result of the post-hoc analysis is presented in Table~\ref{tb:pairwise}. 
The vertical and horizontal dimensions of the Table~\ref{tb:pairwise} show the eleven-by-eleven comparison of the twelve strategies. 
As one can see, we observe various significant differences between the strategies ($p < .05$, marked in bold font).
For example, while we observe a significant difference between the strategies CF-IDF $\times$ Sliding window $\times$ Title and HCF-IDF $\times$ Sliding window $\times$ All ($t(122)=4.77$, $p = .00$), there is no significant difference between the strategies CF-IDF $\times$ Exponential decay $\times$ Title and LDA $\times$ Sliding window $\times$ Title ($t(122)=2.43$, n.s., $p = .41$).
\begin{table}[!ht]
\centering
\small
\caption{Rankscores of the strategies in decreasing order. M and SD denote mean and standard deviation, respectively.}
\begin{tabular}{|c|l|l|l||c|} \hline
		\multirow{2}{*}{} & \multicolumn{3}{c||}{\textbf{Strategy}} & \multicolumn{1}{c|}{\textbf{Rankscore}} \\ \cline{2-5}
		& \multicolumn{1}{c|}{\breakcell{\textbf{Profiling}\\ \textbf{Method}}} & \multicolumn{1}{c|}{\breakcell{\textbf{Decay}\\ \textbf{Function}}} & \multicolumn{1}{c||}{\breakcell{\remove{\textbf{Doc-}\\ \textbf{ment} \\ }\textbf{Con-} \\ \textbf{tent}}} & \multicolumn{1}{c|}{\textbf{M (SD)}} \\  \hline\hline
		1. & CF-IDF & Sliding window & All & .59 (.33)\\  \hline
		2. & HCF-IDF & Sliding window & All & .56 (.34)\\  \hline
		3. & HCF-IDF & Sliding window & Title & .55 (.33)\\  \hline
		4. & HCF-IDF & Exponential decay & Title & .52 (.30)\\  \hline
		5. & CF-IDF & Exponential decay & All & .51 (.32)\\  \hline
		6. & HCF-IDF & Exponential decay & All & .49 (.30)\\  \hline
		7. & CF-IDF & Exponential decay & Title & .41 (.29)\\  \hline
		8. & CF-IDF & Sliding window & Title & .39 (.27)\\  \hline
		9. & LDA & Exponential decay & Title & .35 (.31)\\  \hline
		10. & LDA & Sliding window & Title & .33 (.31)\\  \hline
		11. & LDA & Exponential decay & All & .32 (.30)\\  \hline
		12. & LDA & Sliding window & All & .27 (.33)\\  \hline
\end{tabular}
\label{tb:single-all}
\end{table}
%
\begin{table*}[!ht]
\centering
\caption{Post-hoc analysis with pairwise p-values over the twelve strategies using Shaffer's MSRB procedure. The p-values are marked in bold font if $p < .05$, which indicates a significant difference between the two strategies. Strategies are sorted by rankscores as shown in Table~\ref{tb:single-all}.}
\small
\begin{tabular}{|c|l|l|l||c|c|c|c|c|c|c|c|c|c|c|} \cline{5-15}
		\multicolumn{4}{l|}{\multirow{4}{*}{}} & \rot{All} & \rot{Title} & \rot{Title} & \rot{All} & \rot{All} & \rot{Title} & \rot{Title} & \rot{Title} & \rot{Title} & \rot{All} & \rot{All} \\  \cline{5-15}
		\multicolumn{4}{c|}{} & \rot{\breakcell{Sliding\\window}} & \rot{\breakcell{Sliding\\window}} & \rot{\breakcell{Exponen-\\tial decay}} & \rot{\breakcell{Exponen-\\tial decay}} & \rot{\breakcell{Exponen-\\tial decay}} & \rot{\breakcell{Exponen-\\tial decay}} & \rot{\breakcell{Sliding\\window}} & \rot{\breakcell{Exponen-\\tial decay}} & \rot{\breakcell{Sliding\\window}} & \rot{\breakcell{Exponen-\\tial decay}} & \rot{\breakcell{Sliding\\window}} \\  \cline{5-15}
    \multicolumn{4}{c|}{} & \rot{HCF-IDF} & \rot{HCF-IDF} & \rot{HCF-IDF} & \rot{CF-IDF} & \rot{HCF-IDF} & \rot{CF-IDF} & \rot{CF-IDF} & \rot{LDA} & \rot{LDA} & \rot{LDA} & \rot{LDA} \\  \cline{5-15}
		\multicolumn{4}{c|}{} & 2. & 3. & 4. & 5. & 6. & 7. & 8. & 9. & 10. & 11. & 12. \\ \hhline{----*{10}{|=}|=|}
		1. & CF-IDF & Sliding window & All & .99 & .97 & .72 & .22 & .12 & \textbf{.00} & \textbf{.00} & \textbf{.00} & \textbf{.00} & \textbf{.00} & \textbf{.00} \\  \hline
    2. & HCF-IDF & Sliding window & All & \cellcolor[HTML]{9B9B9B} & .99 & .99 & .99 & .99 & \textbf{.00} & \textbf{.00} & \textbf{.00} & \textbf{.00} & \textbf{.00} & \textbf{.00} \\  \hline
		3. & HCF-IDF & Sliding window & Title & \cellcolor[HTML]{9B9B9B} & \cellcolor[HTML]{9B9B9B} & .99 & .99 & .99 & \textbf{.00} & \textbf{.00} & \textbf{.00} & \textbf{.00} & \textbf{.00} & \textbf{.00} \\  \hline
		4. & HCF-IDF & Exponential decay & Title & \cellcolor[HTML]{9B9B9B} & \cellcolor[HTML]{9B9B9B} & \cellcolor[HTML]{9B9B9B} & .99 & .99 & \textbf{.01} & \textbf{.00} & \textbf{.00} & \textbf{.00} & \textbf{.00} & \textbf{.00} \\  \hline
		5. & CF-IDF & Exponential decay & All & \cellcolor[HTML]{9B9B9B} & \cellcolor[HTML]{9B9B9B} & \cellcolor[HTML]{9B9B9B} & \cellcolor[HTML]{9B9B9B} & .99 & \textbf{.04} & \textbf{.00} & \textbf{.00} & \textbf{.00} & \textbf{.00} & \textbf{.00} \\  \hline
		6. & HCF-IDF & Exponential decay & All & \cellcolor[HTML]{9B9B9B} & \cellcolor[HTML]{9B9B9B} & \cellcolor[HTML]{9B9B9B} & \cellcolor[HTML]{9B9B9B} & \cellcolor[HTML]{9B9B9B} & .12 & \textbf{.02} & \textbf{.00} & \textbf{.00} & \textbf{.00} & \textbf{.00} \\  \hline
		7. & CF-IDF & Exponential decay & Title & \cellcolor[HTML]{9B9B9B} & \cellcolor[HTML]{9B9B9B} & \cellcolor[HTML]{9B9B9B} & \cellcolor[HTML]{9B9B9B}& \cellcolor[HTML]{9B9B9B} & \cellcolor[HTML]{9B9B9B} & .99 & .99 & .41 & .28 & \textbf{.01} \\  \hline
		8. & CF-IDF & Sliding window & Title & \cellcolor[HTML]{9B9B9B} & \cellcolor[HTML]{9B9B9B} & \cellcolor[HTML]{9B9B9B} & \cellcolor[HTML]{9B9B9B} & \cellcolor[HTML]{9B9B9B} & \cellcolor[HTML]{9B9B9B} & \cellcolor[HTML]{9B9B9B} & .99 & .84 & .61 & \textbf{.03} \\  \hline
    9. & LDA & Exponential decay & Title & \cellcolor[HTML]{9B9B9B} & \cellcolor[HTML]{9B9B9B} & \cellcolor[HTML]{9B9B9B} & \cellcolor[HTML]{9B9B9B} & \cellcolor[HTML]{9B9B9B} & \cellcolor[HTML]{9B9B9B} & \cellcolor[HTML]{9B9B9B} & \cellcolor[HTML]{9B9B9B} & .99 & .99 & .72 \\  \hline
		10. & LDA & Sliding window & Title & \cellcolor[HTML]{9B9B9B} & \cellcolor[HTML]{9B9B9B} & \cellcolor[HTML]{9B9B9B} & \cellcolor[HTML]{9B9B9B} & \cellcolor[HTML]{9B9B9B} & \cellcolor[HTML]{9B9B9B} & \cellcolor[HTML]{9B9B9B} & \cellcolor[HTML]{9B9B9B} & \cellcolor[HTML]{9B9B9B} & .99 & .99 \\  \hline
		11. & LDA & Exponential decay & All & \cellcolor[HTML]{9B9B9B} & \cellcolor[HTML]{9B9B9B} & \cellcolor[HTML]{9B9B9B} & \cellcolor[HTML]{9B9B9B} & \cellcolor[HTML]{9B9B9B} & \cellcolor[HTML]{9B9B9B} & \cellcolor[HTML]{9B9B9B} & \cellcolor[HTML]{9B9B9B} & \cellcolor[HTML]{9B9B9B} & \cellcolor[HTML]{9B9B9B} & .88 \\  \hline
\end{tabular}
\label{tb:pairwise}
\end{table*}
\begin{table}[!ht]
\centering
\small
\caption{Three-way repeated-measure ANOVA with Greenhouse-Geisser correction with F-ratio, effect size $\eta^2$, and p-value.}
\begin{tabular}{|l|r|r|r|} \hline
    \multicolumn{1}{|c|}{\textbf{Factor}} & \multicolumn{1}{c|}{\textbf{F}} & \multicolumn{1}{c|}{\textbf{$\eta^2$}} & \multicolumn{1}{c|}{\textbf{p}} \\  \hline\hline
		\textit{Profiling Method} & 58.40 & .48 & \textbf{.00} \\  \hline
    \textit{Decay Function} & 1.17 & .01 & .28 \\  \hline
    \textit{Document Content} & 5.18 & .04 & \textbf{.02} \\  \hline
    \textit{Profiling Method} $\times$ \textit{Decay Function} & 4.63 & .04 & \textbf{.01} \\  \hline
    \textit{Profiling Method} $\times$ \textit{Document Content} & 17.09 & .14 & \textbf{.00} \\  \hline
    \textit{Decay Function} $\times$ \textit{Document Content} & 4.69 & .04 & \textbf{.03} \\  \hline
		\breakcell{\textit{Profiling Method} $\times$ \textit{Decay Function} $\times$ \\ \textit{Document Content}} & 3.35 & .03 & \textbf{.04} \\  \hline
\end{tabular}
\label{tb:three-way-anova}
\end{table}

\textbf{Difference in experiment factors.}
Subsequently, we analyze the results with respect to each experimental factor. 
To this end, we first apply Mendoza's test~\cite{mendoza:1980} to check for violations of sphericity against the factors. 
Mendoza's test is an extension of Mauchly's test to adopt to multi-way repeated-measure ANOVA. 
It shows significances with the global ($\chi^2(65) = 435.90$, $p = .00$) and the factors \textit{Profiling Method} ($\chi^2(2) = 12.21$, $p = .00$), \textit{Profiling Method} $\times$ \textit{Decay Function} ($\chi^2(2) = 20.02$, $p = .00$), and \textit{Profiling Method} $\times$ \textit{Document Content} ($\chi^2(2) = 8.61$, $p = .01$). 
Subsequently, we run a three-way repeated-measure ANOVA with a Greenhouse-Geisser correction of $\epsilon = .60$ for the global and $\epsilon = .91$ for the factors \textit{Profiling Method}, $\epsilon = .87$ for \textit{Profiling Method} $\times$ \textit{Decay Function}, and $\epsilon = .93$ for \textit{Profiling Method} $\times$ \textit{Document Content}. 
Table~\ref{tb:three-way-anova} shows the results of the ANOVA with F-ratio, effect size $\eta^2$, and p-value. 
The effect size is small when $\eta^2 > .02$, medium when $\eta^2 > .13$, and large when $\eta^2 > .26$. 
The analyses reveal significant differences in all three factors and their contributions except the factor \textit{Decay Function}. 
For all factors with significant differences, we apply again a post-hoc analysis using Shaffer's MSRB procedure with respect to each factor. 
In terms of the factor \textit{Profiling Method}, the post-hoc analysis reveals significant differences between all pairs of HCF-IDF, CF-IDF, and LDA (details of the post-hoc analysis are omitted for the reasons of brevity and documented in our TR~\cite{appendix}). 
Although the strategy CF-IDF $\times$ Sliding window $\times$ All performs best as shown in Table~\ref{tb:single-all}, the best \textit{Profiling Method} is HCF-IDF as it performs under all other factors better than CF-IDF and LDA. 
Regarding the factor \textit{Document Content}, ``All'' outperforms ``Title'' ($F(1,122)=5.18$, $p=.02$). 
%
Regarding the factor \textit{Profiling Method} $\times$ \textit{Decay Function}, the result suggests that the strategies with the Exponential decay function perform better than those with the Sliding window function when LDA is employed.  
In addition, there are significant differences among the three profiling methods when a decay function is fixed. 
In both decay functions, HCF-IDF performs best, followed by CF-IDF, and LDA. 
Referring to the factor \textit{Profiling Method} $\times$ \textit{Document Content}, the result indicates that All is a better choice than Title, when CF-IDF is employed. 
In profiling methods HCF-IDF and LDA, the factor \textit{Document Content} makes no significant difference. 
It indicates that HCF-IDF does perform well when only titles of candidate items are available.  
In addition there are significant differences among the profiling methods when a choice of \textit{Document Content} is fixed. 
In those cases, HCF-IDF always outperforms others.  
In terms of the factor \textit{Decay Function} $\times$ \textit{Document Content}, All is a better choice than Title, when Sliding window is used.

\textbf{Correlation of recommendation performance with the number of tweets, the number of concepts, the number of concepts per tweet, and the percentage of tweets containing at least one concept.}
We computed Pearson's $r$ and Kendall's $\tau$ between the users' mean rankscores and each of the number of tweets, concepts, concepts per tweet and the percentage of tweets containing at least one concept. 
A correlation may show a dependency that could influence the recommendation performance. 
The results show no significant correlation: 
As stated in Section~\ref{sec:dataset}, a participant has published on average $1096.82$ tweets (SD: $1048.46$). 
There is no significant correlation with the rankscores ($r(121) = .04$, n.s., $p = .62$ and $\tau = .00$, n.s., $p = .98$). 
Referring to the number of concepts, on average $1,214.93$ concepts (SD: $1181.43$) are contained in a participant's Twitter stream. 
The correlation coefficients are non-significant ($r(121) = .05$, n.s., $p = .60$ and $\tau = -.01$, n.s., $p = .94$). 
Regarding the number of concepts per tweet, a participant's tweet contains on average $1.07$ concepts (SD: $0.31$) with again no significant correlation to the rankscores ($r(121) = -.05$, n.s., $p = .59$ and $\tau = -.02$, n.s., $p = .71$). 
Regarding the tweets that contribute in computing the user profiles for the methods with CF-IDF and HCF-IDF, we calculate the percentage of the number of tweets containing at least one concept and the number of tweets for each user. 
On average, $62.24$\% of the tweets (SD: $13.55$) that a participant has published contain at least one concept, with no significant correlation ($r(121) = -.04$, n.s., $p = .67$ and $\tau = -.03$, n.s., $p = .73$)
\subsection{Questionnaire Feedback}
\label{sec:qualititative-feedback}
At the end of the experiment, the participants were asked to rate: ``How easy it was to make the decisions whether a recommended publication is interesting".
Using a 5-point Likert scale, where values between 1 and 5 refer to very difficult to very easy, the result is fairly high with an average of $3.68$ (SD: $0.88$). 
Regarding question ``Whether the participants noticed a difference among the twelve strategies", the result is similarly high with an average of $3.46$ (SD: $1.20$). 
In the free text feedback, one participant denoted that the recommender system failed to pick up his primary field despite having tweeted about that field. 
Apart from this, we received many positive comments (e\,.g., interesting, useful). 
%
\section{Discussion}
\label{sec:discussion}
The strategies with HCF-IDF perform almost equally well compared to the best performing strategy CF-IDF $\times$ Sliding window $\times$ All. 
There is no significant difference between them as described in Table~\ref{tb:pairwise}. 
The strong advantage of HCF-IDF is that it reaches its performance already when using only the titles of the scientific publications.
The reason is that spreading activation over the hierarchical knowledge base used in HCF-IDF successfully reveals concepts that are not explicitly mentioned in the texts. 
CF-IDF works well when full-texts are available. 
Referring to LDA, the recommendation performance of the strategies with LDA is overall low, even if full-texts are available. 
A possible reason is that LDA cannot construct accurate user profiles because of the shortness and sparseness of social media items. 
Without accurate user profiles it is impossible to make good recommendations, even if full-texts are available. 
In fact, a slight correlation between the rankscores of LDA and the number of tweets is observed~\cite{appendix}. 
It indicates that participants with more tweets receive better recommendations. 
Please note as documented in~\cite{appendix}, rankscores are almost exact same values with Precision@5 and nDCG. 
Although rankscores are slightly different with MAP and MRR, the order of performance of strategies are almost identical. 
Thus, the arguments described in this paper do not be influenced by differences among those evaluation metrics. 
Our dataset covers scientific publications in the broader field of economics.
Thus, although the dataset is obtained from a portal of economics literature, it contains scientific publications from various fields including, e\,.g., social sciences, political sciences, and information sciences.
In the experiment, $31$ of $123$ participants do not have a major in economics. 
We have conducted an ANOVA test to identify whether the recommendation performance is significantly different for participants from economics and those not in economics.
The result shows that majors make no significant difference ($F(1,121) = 0.01$, n.s., $p = .94$). 
Thus, we assume that our approach may be transferred to other domains. 
%
Furthermore, there are a lot of domain-specific hierarchical knowledge bases in other domains freely available such as Medical Subject Headings (MeSH) for medicine and ACM Computing Classification System (ACM CCS) for computer science. 
An overview of freely available hierarchical knowledge bases is maintained by the W3C as cited in the introduction. 
The knowledge bases are of similar structure to the STW used in this paper. 
They are of high quality as they are manually crafted by domain experts. 
Therefore, HCF-IDF can be easily applied to other fields. 
Our approach could be integrated with other social media platforms (e\,.g., Facebook, LinkedIn), where users generate short and sparse texts. 
In addition, HCF-IDF is robust against the number of tweets a user published, because there is no correlation between the number of tweets and the rankscores of the strategies with HCF-IDF. 

%

Our results may potentially be influenced by the amount of time that each participant spent for evaluating the $5 \times 12 = 60$ recommended publications by the twelve strategies in the experiment. 
However, they spent on average $517.54$ seconds (SD:~$376.72$) to complete the evaluation of the $60$ recommendations. 
In addition, we randomized the order of the strategies presented to the participants to counterbalance any influence on the order of the strategies. 
Thus, we think that our results are not influenced by it. 
Another potential threat to the validity of our results could be the procedure how we recruited the participants.
We believe that the risk is low since we collected enough participants regarding each demographic factor (as shown in Section~\ref{sec:participants}). 
%
Regarding the demographic factors, we found significant differences only for the participants' highest academic degree and participants' gender (details are documented in the TR~\cite{appendix}).
However, they do not affect the order of the recommendation performance of the different strategies. 
%
%
\remove{Thus, we also assume that if users have published more than the maximum of 3200 tweets obtainable from the Twitter API, this doesn't have a negative impact.} 
\section{Conclusions}
\label{sec:conclusion}

This paper contributes to content-based recommender systems for scientific publications based on user profiles extracted from social media platforms.
We have constructed twelve different recommendation strategies along three factors, namely profiling method, decay function, and document content. 
The online experiment revealed that titles of scientific publications are sufficient to achieve competitive recommendation results when employing the profiling method HCF-IDF.
Thus, the spreading activation over the hierarchical knowledge base enables HCF-IDF to extract a sufficient number of concepts from titles to compute competitive recommendations.
This is an important result as full-texts are not always available, e.\,g., due to legal reasons.
%

\remove{
The results of this experiment can be deployed as a separate service for recommending scholarly content.
In addition, the results suggest that if a background taxonomy is available the HCF-IDF $\times$ Sliding window $\times$ Title strategy can be used to generate recommendations.

In practice, it is not easy to obtain and automatically process full-texts of scientific publications, e.\,g., due to legal issues. 
In addition, HCF-IDF saves a lot of computational cost and storage capacity when applied to title data only. 
They can also be employed to enhance the services of scientific publication search engines, extending a user's search query with extracted user profiles. 
}
 
\paragraph*{Acknowledgement}
This research was co-financed by the EU H2020 project MOVING (\url{http://www.moving-project.eu/}) under contract no 693092.
We like to thank the anonymous participants of our study to support this research.

\bibliographystyle{plain}
{
\bibliography{ref}
}

\clearpage

\appendix
\section{Three-way repeated ANOVA for rankscores}
We describe the details of the three-way repeated ANOVA for rankscores conducted in Section~\ref{sec:quantitative-analyses}. 
Specifically, we provide the analyses with respect to each factor that shows a significant difference as documented in Table~\ref{tb:three-way-anova}. 
For all statistical analyses, we use $\alpha = .05$ as significance level. 
Regarding the effect size $\eta^2$, the effect size is small when $\eta^2 > .02$, medium when $\eta^2 > .13$, and large when $\eta^2 > .26$.
In terms of the effect size $d$ measured by Cohen's $d$, the effect size is interpreted small when $d = .20$, medium when $d = .50$, and large when $d = .80$. 
\paragraph*{\textbf{The factor \textit{Profiling Method}}}
Tables~\ref{tb:post-i}(a), (b) and (c) show the rankscores, the post-hoc analysis for the factor \textit{Profiling Method}, and the effect size, respectively. 
Table~\ref{tb:post-i}(a) presents the means and standard deviations of the three profiling methods. 
Table~\ref{tb:post-i}(b) shows p-values of each pair. 
Since Table~\ref{tb:three-way-anova} shows that this factor has the largest effect size, we further compute the effect size using Cohen's d for each pair shown in Table~\ref{tb:post-i}(c). 
\begin{table}[!h]
\small
\centering
\caption{Rankscores, Post-hoc analysis for the factor \textit{Profiling Method} using Shaffer's MSRB procedure, and effect size.}
\begin{tabular}{|c|c|c|}
		\multicolumn{3}{c}{\textbf{a) Rankscores}} \\  \hline
    \multicolumn{1}{|c|}{\textbf{Choice}} & \multicolumn{1}{c|}{\textbf{M}} & \multicolumn{1}{c|}{\textbf{SD}} \\  \hline\hline
    HCF-IDF & .53 & .32 \\  \hline
		CF-IDF & .48 & .31 \\  \hline
	  LDA & .32 & .31 \\  \hline
		\multicolumn{3}{c}{\textbf{b) Post-hoc analysis p-values}} \\  \hline
     & HCF-IDF & LDA \\  \hline \hline
	  CF-IDF & \textbf{.00} & \textbf{.00}  \\  \hline
    HCF-IDF & \cellcolor[HTML]{9B9B9B} & \textbf{.00} \\  \hline
		\multicolumn{3}{c}{\textbf{c) Effect size using Cohen's d}} \\  \hline
     & HCF-IDF & LDA \\  \hline \hline
	  CF-IDF & .17 & .50 \\  \hline
    HCF-IDF & \cellcolor[HTML]{9B9B9B} & .67 \\  \hline
\end{tabular}
\label{tb:post-i}
\end{table}
\paragraph*{\textbf{The factor \textit{Document Content}}}
Table~\ref{tb:post-iii} shows the post-hoc analysis for the factor \textit{Document Content}. 
The result shows that the recommender systems perform better when using both titles and full-texts. 
\begin{table}[!ht]
\small
\centering
\caption{Rankscores and Post-hoc analysis for the factor \textit{Document Content} using Shaffer's MSRB procedure.}
\begin{tabular}{|c|c|c|}
		\multicolumn{3}{c}{\textbf{a) Rankscores}} \\  \hline
    \multicolumn{1}{|c|}{\textbf{Choice}} & \multicolumn{1}{c|}{\textbf{M}} & \multicolumn{1}{c|}{\textbf{SD}} \\  \hline\hline
    All & .46 & .21 \\  \hline
		Title & .43 & .20 \\  \hline
		\multicolumn{3}{c}{\textbf{b) Post-hoc analysis p-values}} \\  \hline
     & \multicolumn{2}{c|}{Title} \\  \hline \hline
	  All & \multicolumn{2}{c|}{\textbf{.02}}  \\  \hline
\end{tabular}
\label{tb:post-iii}
\end{table}
\paragraph*{\textbf{The factor \textit{Profiling Method} $\times$ \textit{Decay Function}}}
Table~\ref{tb:anova-i-ii} shows the results of ANOVA regarding the factor \textit{Profiling Method} when a choice of the factor \textit{Decay Function} is fixed and vice versa. 
Mendoza's test found a violation of sphericity in the factor \textit{Profiling Method} when Sliding window is used ($\chi^2(2) = 9.26$, $p=.01$) and Exponential decay is used ($\chi^2(2) = 11.16$, $p=.00$). 
Thus, we run a one-way repeated-measure ANOVA with Greenhouse-Geisser correction of $\eta = .93$ for the first row in Table~\ref{tb:anova-i-iii} and $\eta = .92$ for the second row in Table~\ref{tb:anova-i-iii}. 
We observe significant differences when a choice of the factor \textit{Decay Function} is fixed and when LDA is employed. 
The post-hoc analyses of them are shown in Table~\ref{tb:post-i-sw}, Table~\ref{tb:post-i-e}, and Table~\ref{tb:post-ii-t}, respectively. 
In Table~\ref{tb:post-i-sw} and Table~\ref{tb:post-i-e}, a choice of the factor \textit{Decay Function} is fixed. 
The results demonstrate that HCF-IDF performs best, followed by CF-IDF and LDA. 
Table~\ref{tb:post-ii-t} shows the post-hoc analysis of the factor \textit{Decay Function} when LDA is employed. 
It indicates Exponential decay performs better than Sliding window for LDA. 
\begin{table}[!ht]
\centering
\small
\caption{ANOVA for \textit{Profiling Method} $\times$ \textit{Decay Function} interaction}
\begin{tabular}{|l|r|r|r|} \hline
    \multicolumn{1}{|c|}{\textbf{Factor}} & \multicolumn{1}{c|}{\textbf{F}} & \multicolumn{1}{c|}{\textbf{$\eta^2$}} & \multicolumn{1}{c|}{\textbf{p}} \\  \hline\hline
		\textit{Profiling Method} at Sliding window & 52.71 & .43 & \textbf{.00} \\  \hline
    \textit{Profiling Method} at Exponential decay & 26.89 & .22 & \textbf{.00} \\  \hline
    \textit{Decay Function} at CF-IDF & 3.69 & .03 & .06 \\  \hline
    \textit{Decay Function} at HCF-IDF & 2.33 & .02 & .12 \\  \hline
    \textit{Decay Function} at LDA & 5.26 & .04 & \textbf{.02} \\  \hline
\end{tabular}
\label{tb:anova-i-ii}
\end{table}
\begin{table}[!ht]
\small
\centering
\caption{Rankscores and Post-hoc analysis for the factor \textit{Profiling Method} at Sliding window using Shaffer's MSRB procedure.}
\begin{tabular}{|c|c|c|}
    \multicolumn{3}{c}{\textbf{a) Rankscores}} \\  \hline
    \multicolumn{1}{|c|}{\textbf{Choice}} & \multicolumn{1}{c|}{\textbf{M}} & \multicolumn{1}{c|}{\textbf{SD}} \\  \hline\hline
    HCF-IDF & .55 & .33 \\  \hline
		CF-IDF & .49 & .32 \\  \hline
	  LDA & .30 & .32 \\  \hline
    \multicolumn{3}{c}{\textbf{b) Post-hoc analysis p-values}} \\  \hline
    & HCF-IDF & LDA \\  \hline \hline
	  CF-IDF & \textbf{.01} & \textbf{.00}  \\  \hline
    HCF-IDF & \cellcolor[HTML]{9B9B9B} & \textbf{.00} \\  \hline
\end{tabular}
\label{tb:post-i-sw}
\end{table}
\begin{table}[!ht]
\small
\centering
\caption{Rankscores and Post-hoc analysis for the factor \textit{Profiling Method} at Exponential decay using Shaffer's MSRB procedure.}
\begin{tabular}{|c|c|c|}
		\multicolumn{3}{c}{\textbf{a) Rankscores}} \\  \hline
    \multicolumn{1}{|c|}{\textbf{Choice}} & \multicolumn{1}{c|}{\textbf{M}} & \multicolumn{1}{c|}{\textbf{SD}} \\  \hline\hline
    HCF-IDF & .51 & .30 \\  \hline
		CF-IDF & .46 & .31 \\  \hline
	  LDA & .34 & .31 \\  \hline
		\multicolumn{3}{c}{\textbf{b) Post-hoc analysis p-values}} \\  \hline
        & HCF-IDF & LDA \\  \hline \hline
	  CF-IDF & \textbf{.02} & \textbf{.00}  \\  \hline
    HCF-IDF & \cellcolor[HTML]{9B9B9B} & \textbf{.00} \\  \hline
\end{tabular}
\label{tb:post-i-e}
\end{table}
\begin{table}[!ht]
\centering
\small
\caption{Rankscores and Post-hoc analysis for the factor \textit{Decay Function} at LDA using Shaffer's MSRB procedure.}
\begin{tabular}{|c|c|c|}
		\multicolumn{3}{c}{\textbf{a) Rankscores}} \\  \hline
    \multicolumn{1}{|c|}{\textbf{Choice}} & \multicolumn{1}{c|}{\textbf{M}} & \multicolumn{1}{c|}{\textbf{SD}} \\  \hline\hline
		Exponential decay & .34 & .31 \\  \hline
    Sliding window & .30 & .32 \\  \hline
		\multicolumn{3}{c}{\textbf{b) Post-hoc analysis p-value}} \\  \hline
        & \multicolumn{2}{c|}{Exponential decay} \\  \hline \hline
	  Sliding window & \multicolumn{2}{c|}{\textbf{.02}} \\  \hline
\end{tabular}
\label{tb:post-ii-t}
\end{table}
\paragraph*{\textbf{The factor \textit{Profiling Method} $\times$ \textit{Document Content}}}
Table~\ref{tb:anova-i-iii} shows the results of ANOVA regarding the factor \textit{Profiling Method} when a choice of the factor \textit{Document Content} is fixed and vice versa. 
We observe there is a significant difference when a choice of the factor \textit{Document Content} is fixed and CF-IDF is employed. 
Mendoza's test found a violation of sphericity in the factor \textit{Profiling Method} when All (i\,.e., titles and full-texts) is used ($\chi^2(2) = 25.24$, $p=.00$). 
Thus, we run a one-way repeated-measure ANOVA with Greenhouse-Geisser correction of $\eta = .84$ for the second row in Table~\ref{tb:anova-i-iii}. 
Table~\ref{tb:post-i-t} presents the post-hoc analysis when Title is selected for the factor \textit{Document Profiling}. 
We see that HCF-IDF outperforms others with significant differences. 
On the other hand, Table~\ref{tb:post-i-f} shows the post-hoc analysis when All is chosen for the factor \textit{Document Profiling}. 
There is no significant difference between CF-IDF and HCF-IDF. 
Table~\ref{tb:post-iii-c} shows the post-hoc analysis of the factor \textit{Document Content} when CF-IDF is employed. 
It indicates that the strategies with CF-IDF and All significantly outperforms those with CF-IDF and Title. 
\begin{table}[!ht]
\small
\centering
\caption{ANOVA for \textit{Profiling Method} $\times$ \textit{Document Content} interaction}
\begin{tabular}{|l|r|r|r|} \hline
    \multicolumn{1}{|c|}{\textbf{Factor}} & \multicolumn{1}{c|}{\textbf{F}} & \multicolumn{1}{c|}{\textbf{$\eta^2$}} & \multicolumn{1}{c|}{\textbf{p}} \\  \hline\hline
		\textit{Profiling Method} at Title & 26.15 & .21 & \textbf{.00} \\  \hline
    \textit{Profiling Method} at All & 55.28 & .45 & \textbf{.00} \\  \hline
    \textit{Document Content} at CF-IDF & 32.95 & .27 & \textbf{.00} \\  \hline
    \textit{Document Content} at HCF-IDF & 0.43 & .00 & .51 \\  \hline
    \textit{Document Content} at LDA & 2.06 & .02 & .15 \\  \hline
\end{tabular}
\label{tb:anova-i-iii}
\end{table}
\begin{table}[!ht]
\centering
\small
\caption{Rankscores and Post-hoc analysis for the factor \textit{Profiling Method} at Title using Shaffer's MSRB procedure.}
\begin{tabular}{|c|c|c|}
		\multicolumn{3}{c}{\textbf{a) Rankscores}} \\  \hline
    \multicolumn{1}{|c|}{\textbf{Choice}} & \multicolumn{1}{c|}{\textbf{M}} & \multicolumn{1}{c|}{\textbf{SD}} \\  \hline\hline
    HCF-IDF & .54 & .31 \\  \hline
		CF-IDF & .40 & .28 \\  \hline
	  LDA & .34 & .31 \\  \hline
		\multicolumn{3}{c}{\textbf{b) Post-hoc analysis p-values}} \\  \hline
        & HCF-IDF & LDA \\  \hline \hline
	  CF-IDF & \textbf{.00} & \textbf{.04}  \\  \hline
    HCF-IDF & \cellcolor[HTML]{9B9B9B} & \textbf{.00} \\  \hline
\end{tabular}
\label{tb:post-i-t}
\end{table}
\begin{table}[!ht]
\centering
\small
\caption{Rankscores and Post-hoc analysis for the factor \textit{Profiling Method} at All using Shaffer's MSRB procedure.}
\begin{tabular}{|c|c|c|}
		\multicolumn{3}{c}{\textbf{a) Rankscores}} \\  \hline
    \multicolumn{1}{|c|}{\textbf{Choice}} & \multicolumn{1}{c|}{\textbf{M}} & \multicolumn{1}{c|}{\textbf{SD}} \\  \hline\hline
		CF-IDF & .55 & .33 \\  \hline
    HCF-IDF & .53 & .32 \\  \hline
	  LDA & .30 & .32 \\  \hline
		\multicolumn{3}{c}{\textbf{b) Post-hoc analysis p-values}} \\  \hline
        & HCF-IDF & LDA \\  \hline \hline
	  CF-IDF & .20 & \textbf{.00}  \\  \hline
    HCF-IDF & \cellcolor[HTML]{9B9B9B} & \textbf{.00} \\  \hline
\end{tabular}
\label{tb:post-i-f}
\end{table}
\begin{table}[!ht]
\centering
\small
\caption{Rankscores and Post-hoc analysis for the factor \textit{Document Content} at CF-IDF using Shaffer's MSRB procedure.}
\begin{tabular}{|c|c|c|}
		\multicolumn{3}{c}{\textbf{a) Rankscores}} \\  \hline
    \multicolumn{1}{|c|}{\textbf{Choice}} & \multicolumn{1}{c|}{\textbf{M}} & \multicolumn{1}{c|}{\textbf{SD}} \\  \hline\hline
		All & .55 & .33 \\  \hline
    Title & .40 & .28 \\  \hline
		\multicolumn{3}{c}{\textbf{b) Post-hoc analysis p-value}} \\  \hline
        & \multicolumn{2}{c|}{All} \\  \hline \hline
	  Title & \multicolumn{2}{c|}{\textbf{.00}} \\  \hline
\end{tabular}
\label{tb:post-iii-c}
\end{table}
\paragraph*{\textbf{The factor \textit{Decay Function} $\times$ \textit{Document Content}}}
Table~\ref{tb:anova-ii-iii} shows the results of ANOVA regarding the factor \textit{Decay Function} when a choice of the factor \textit{Document Content} is fixed and vice versa. 
According to Table~\ref{tb:anova-ii-iii}, there is a significant difference among the factor \textit{Document Content}, when Sliding window is used. 
The rankscores and post-hoc analysis of it are shown in Tables~\ref{tb:post-iii-s}(a) and (b). 
It indicates that All significantly enhances the performance of the recommender system when Sliding window is used. 
\begin{table}[!ht]
\centering
\small
\caption{ANOVA for \textit{Decay Function} $\times$ \textit{Document Content} interaction}
\begin{tabular}{|l|r|r|r|} \hline
    \multicolumn{1}{|c|}{\textbf{Factor}} & \multicolumn{1}{c|}{\textbf{F}} & \multicolumn{1}{c|}{\textbf{$\eta^2$}} & \multicolumn{1}{c|}{\textbf{p}} \\  \hline\hline
		{\textit{Decay Function}} at Title & 0.04 & .00 & .85 \\  \hline
    {\textit{Decay Function}} at All & 3.16 & .03 & .08 \\  \hline
    \textit{Document Content} at Sliding window & 9.44 & .08 & \textbf{.00} \\  \hline
    \textit{Document Content} at Exponential decay & 0.56 & .00 & .46 \\  \hline
\end{tabular}
\label{tb:anova-ii-iii}
\end{table}
\begin{table}[!ht]
\centering
\small
\caption{Rankscores and Post-hoc analysis for the factor \textit{Document Content} at Sliding window using Shaffer's MSRB procedure.}
\begin{tabular}{|c|c|c|}
		\multicolumn{3}{c}{\textbf{a) Rankscores}} \\  \hline
    \multicolumn{1}{|c|}{\textbf{Choice}} & \multicolumn{1}{c|}{\textbf{M}} & \multicolumn{1}{c|}{\textbf{SD}} \\  \hline\hline
		All & .48 & .36 \\  \hline
    Title & .42 & .32 \\  \hline
		\multicolumn{3}{c}{\textbf{b) Post-hoc analysis p-value}} \\  \hline
        & \multicolumn{2}{c|}{All} \\  \hline \hline
	  Title & \multicolumn{2}{c|}{\textbf{.00}} \\  \hline
\end{tabular}
\label{tb:post-iii-s}
\end{table}
\remove{
\section{Precision at each rank}
In addition, we compute precisions at each rank for the best three and worst three strategies described in Table~\ref{tb:single-all}. Table~\ref{tb:precision-rank} shows the results. 
For the best three strategies, the precision follow the almost uniform distribution over ranks. Regarding the worst three strategies, specifically T-E-F and T-S-F, the precision declines slightly as rank goes down. 
\begin{table}[!ht]
\centering
\small
\caption{Precision at each rank for the best and worst three strategies}
\begin{tabular}{|c|c|c|c||c|c|c|} \hline
    \multicolumn{1}{|c|}{} & \multicolumn{1}{c|}{\textbf{C-S-F}} & \multicolumn{1}{c|}{\textbf{H-S-F}} & \multicolumn{1}{c||}{\textbf{H-S-T}} & \multicolumn{1}{c|}{\textbf{T-S-T}} & \multicolumn{1}{c|}{\textbf{T-E-F}} & \multicolumn{1}{c|}{\textbf{T-S-F}} \\  \hline\hline
	  1 & .59 & .52 & .53 & .25 & .38 & .33 \\  \hline
    2 & .61 & .63 & .60 & .33 & .32 & .26 \\  \hline
    3 & .61 & .50 & .54 & .44 & .29 & .24 \\  \hline
		4 & .59 & .60 & .56 & .33 & .28 & .24 \\  \hline
		5 & .55 & .55 & .50 & .34 & .27 & .26 \\  \hline
\end{tabular}
\label{tb:precision-rank}
\end{table}
}
\section{Mean Average Precision}
We look into the recommendation performance by computing Mean Average Precision (MAP). 
Average Precision (AP) is calculated as described in Equation~\ref{eq:ap}. 
\begin{equation}
AP = \frac{1}{|hits|}\sum_{d \in hits}{Precision@rank_{d}}, 
\label{eq:ap}
\end{equation}
where $hits$ and $rank_d$ stand for the set of relevant items and the rank of the item $d$, respectively. 
$|hits|$ is the number of relevant items in the recommendation list. 
$Precision@rank_{d}$ denotes the precision at cut off $rank_{d}$ in the recommendation list. 
Mean Average Precision (MAP) is the mean of the Average Precision scores for each participant. 
In this section, we evaluate the recommendation performance using MAP. 
Particularly, we first compare the twelve different strategies. 
Subsequently, we investigate the difference among the different experiment factors. 
\subsection{Best performing strategy}
Table~\ref{tb:map} shows the Mean Average Precisions (MAP) of the twelve strategies. 
The order of the strategies is almost same with rankscores shown in~Table \ref{tb:single-all}. 
In order to investigate significant differences among strategies, we first apply Mauchly's test and found a violation of sphericity in the strategies ($\chi^2(65) = 353.51$, $p = .00$). 
Subsequently, we run a one-way repeated-measure ANOVA with a Greenhouse-Geisser correction of $\epsilon = .65$. 
It reveals a significant difference of the strategies' MAPs ($F(7.17, 875.15) = 15.59$, $p = .00$). 
To assess the statistical significance of pair-wise differences between the twelve strategies, a post-hoc analysis is performed using Shaffer's MSRB procedure~\cite{shaffer:1986}. 
The result of the post-hoc analysis is presented in Table~\ref{tb:pairwise-map}. 
The vertical and horizontal dimensions of the Table~\ref{tb:pairwise-map} show the eleven-by-eleven comparison of the twelve strategies. 
As one can see, we observe various significant differences between strategies (marked in bold font).
\begin{table}[!ht]
\centering
\small
\caption{Mean Average Precision (MAP) of the strategies in decreasing order.}
\begin{tabular}{|c|l|l|l||c|} \hline
		\multirow{2}{*}{} & \multicolumn{3}{c||}{\textbf{Strategy}} & \multicolumn{1}{c|}{\breakcell{\textbf{MAP}}} \\  \cline{2-5}
		& \multicolumn{1}{c|}{\breakcell{\textbf{Profiling}\\ \textbf{Method}}} & \multicolumn{1}{c|}{\breakcell{\textbf{Decay}\\ \textbf{Function}}} & \multicolumn{1}{c||}{\breakcell{\textbf{Con-} \\ \textbf{tent}}} & \multicolumn{1}{c|}{\textbf{M (SD)}} \\  \hline\hline
		1. & CF-IDF & Sliding window & All & .71 (.32) \\  \hline
		2. & HCF-IDF & Exponential decay & All & .65 (.33) \\  \hline
		3. & HCF-IDF & Exponential decay & Title & .65 (.32) \\  \hline
		4. & CF-IDF & Exponential decay & All & .65 (.35) \\  \hline
		5. & HCF-IDF & Sliding window & Title & .65 (.34) \\  \hline
		6. & HCF-IDF & Sliding window & All & .65 (.34) \\  \hline
		7. & CF-IDF & Exponential decay & Title & .58 (.35) \\  \hline
		8. & CF-IDF & Sliding window & Title & .55 (.34) \\  \hline
		9. & LDA & Exponential decay & All & .47 (.39) \\  \hline
		10. & LDA & Exponential decay & Title & .44 (.34) \\  \hline
		11. & LDA & Sliding window & Title & .43 (.35) \\  \hline
		12. & LDA & Sliding window & All & .40 (.42) \\  \hline
\end{tabular}
\label{tb:map}
\end{table}
%
\begin{table*}[!ht]
\centering
\medskip
\caption{Post-hoc analysis of Mean Average Precision (MAP) with pairwise p-values over the strategies using Shaffer's MSRB procedure. The p-values are marked in bold font if $p < .05$, which indicates a significant difference between the two strategies. Strategies are sorted by Precision@5 as shown in Table~\ref{tb:map}.}
\small
\begin{tabular}{|c|l|l|l||c|c|c|c|c|c|c|c|c|c|c|} \cline{5-15}
		\multicolumn{4}{l|}{\multirow{4}{*}{}} & \rot{All} & \rot{Title} & \rot{All} & \rot{Title} & \rot{All} & \rot{Title} & \rot{Title} & \rot{All} & \rot{Title} & \rot{Title} & \rot{All} \\  \cline{5-15}
		\multicolumn{4}{c|}{} & \rot{\breakcell{Exponen-\\tial decay}} & \rot{\breakcell{Exponen-\\tial decay}} & \rot{\breakcell{Exponen-\\tial decay}} & \rot{\breakcell{Sliding\\window}} & \rot{\breakcell{Sliding\\window}} & \rot{\breakcell{Exponen-\\tial decay}} & \rot{\breakcell{Sliding\\window}} & \rot{\breakcell{Exponen-\\tial decay}} & \rot{\breakcell{Exponen-\\tial decay}} & \rot{\breakcell{Sliding\\window}} & \rot{\breakcell{Sliding\\window}} \\  \cline{5-15}
		\multicolumn{4}{c|}{} & \rot{HCF-IDF} & \rot{HCF-IDF} & \rot{CF-IDF} & \rot{HCF-IDF} & \rot{HCF-IDF} & \rot{CF-IDF} & \rot{CF-IDF} & \rot{LDA} & \rot{LDA} & \rot{LDA} & \rot{LDA} \\  \cline{5-15}
		\multicolumn{4}{c|}{} & 2. & 3. & 4. & 5. & 6. & 7. & 8. & 9. & 10. & 11. & 12. \\ \hhline{----*{10}{|=}|=|}
		1. & CF-IDF & Sliding window & All & .99 & .99 & .99 & .99 & .99 & \textbf{.02} & \textbf{.00} & \textbf{.00} & \textbf{.00} & \textbf{.00} & \textbf{.00} \\  \hline
		2. & HCF-IDF & Exponential decay & All & \cellcolor[HTML]{9B9B9B} & .99 & .99 & .99 & .99 & .99 & .25 & \textbf{.00} & \textbf{.00} & \textbf{.00} & \textbf{.00} \\  \hline
		3. & HCF-IDF & Exponential decay & Title & \cellcolor[HTML]{9B9B9B} & \cellcolor[HTML]{9B9B9B} & .99 & .99 & .99 & .99 & .99 & \textbf{.00} & \textbf{.00} & \textbf{.00} & \textbf{.00} \\  \hline
		4. & CF-IDF & Exponential decay & All & \cellcolor[HTML]{9B9B9B} & \cellcolor[HTML]{9B9B9B} & \cellcolor[HTML]{9B9B9B} & .99 & .99 & .99 & .33 & \textbf{.01} & \textbf{.00} & \textbf{.00} & \textbf{.00} \\  \hline
		5. & HCF-IDF & Sliding window & Title & \cellcolor[HTML]{9B9B9B} & \cellcolor[HTML]{9B9B9B} & \cellcolor[HTML]{9B9B9B} & \cellcolor[HTML]{9B9B9B} & .99 & .99 & .54 & \textbf{.00} & \textbf{.00} & \textbf{.00} & \textbf{.00} \\  \hline
		6. & HCF-IDF & Sliding window & All & \cellcolor[HTML]{9B9B9B} & \cellcolor[HTML]{9B9B9B} & \cellcolor[HTML]{9B9B9B} & \cellcolor[HTML]{9B9B9B} & \cellcolor[HTML]{9B9B9B} & .99 & .60 & \textbf{.00} & \textbf{.00} & \textbf{.00} & \textbf{.00} \\  \hline
    7. & CF-IDF & Exponential decay & Title & \cellcolor[HTML]{9B9B9B} & \cellcolor[HTML]{9B9B9B} & \cellcolor[HTML]{9B9B9B} & \cellcolor[HTML]{9B9B9B} & \cellcolor[HTML]{9B9B9B} & \cellcolor[HTML]{9B9B9B} & .99 & .67 & \textbf{.02} & \textbf{.01} & \textbf{.00} \\  \hline
    8. & CF-IDF & Sliding window & Title & \cellcolor[HTML]{9B9B9B} & \cellcolor[HTML]{9B9B9B} & \cellcolor[HTML]{9B9B9B} & \cellcolor[HTML]{9B9B9B} & \cellcolor[HTML]{9B9B9B} & \cellcolor[HTML]{9B9B9B} & \cellcolor[HTML]{9B9B9B} & .99 & .06 & \textbf{.03} & \textbf{.02} \\  \hline
    9. & LDA & Exponential decay & All & \cellcolor[HTML]{9B9B9B} & \cellcolor[HTML]{9B9B9B} & \cellcolor[HTML]{9B9B9B} & \cellcolor[HTML]{9B9B9B} & \cellcolor[HTML]{9B9B9B} & \cellcolor[HTML]{9B9B9B} & \cellcolor[HTML]{9B9B9B} & \cellcolor[HTML]{9B9B9B} & .99 & .99 & .64 \\  \hline
		10. & LDA & Exponential decay & Title & \cellcolor[HTML]{9B9B9B} & \cellcolor[HTML]{9B9B9B} & \cellcolor[HTML]{9B9B9B} & \cellcolor[HTML]{9B9B9B} & \cellcolor[HTML]{9B9B9B} & \cellcolor[HTML]{9B9B9B} & \cellcolor[HTML]{9B9B9B} & \cellcolor[HTML]{9B9B9B} & \cellcolor[HTML]{9B9B9B} & .99 & .99 \\  \hline
		11. & LDA & Sliding window & Title & \cellcolor[HTML]{9B9B9B} & \cellcolor[HTML]{9B9B9B} & \cellcolor[HTML]{9B9B9B} & \cellcolor[HTML]{9B9B9B} & \cellcolor[HTML]{9B9B9B} & \cellcolor[HTML]{9B9B9B} & \cellcolor[HTML]{9B9B9B} & \cellcolor[HTML]{9B9B9B} & \cellcolor[HTML]{9B9B9B} & \cellcolor[HTML]{9B9B9B} & .99 \\  \hline
\end{tabular}
\label{tb:pairwise-map}
\end{table*}
\subsection{Difference in experiment factors}
Subsequently, we analyze the results of MAPs with respect to each factor. 
First, we apply Mendoza's test~\cite{mendoza:1980} which shows violations of sphericity against the factors \textit{Profiling Method} $\times$ \textit{Decay Function} ($\chi^2(2) = 10.30$, $p = .01$), and \textit{Profiling Method} $\times$ \textit{Document Content} ($\chi^2(2) = 13.18$, $p = .00$). 
Thus, we run three-way repeated-measure ANOVA with a Greenhouse-Geisser correction of $\epsilon = .92$ for the factor \textit{Profiling Method} $\times$ \textit{Decay Function}, and $\epsilon = .91$ for the factor \textit{Profiling Method} $\times$ \textit{Document Content}. 
Table~\ref{tb:three-way-anova-map} shows the results of applying an ANOVA. 
$\eta^2$ indicates the effect size of each factor. 
For all the factors that make significant difference, we conduct a post-hoc analysis using Shaffer's MSRB Procedure. 
\begin{table}[!ht]
\centering
\medskip
\small
\caption{Three-way repeated-measure ANOVA with Greenhouse-Geisser correction with F-ratio, $\eta^2$, and p-value for MAP.}
\begin{tabular}{|l|r|r|r|} \hline
    \multicolumn{1}{|c|}{\textbf{Factor}} & \multicolumn{1}{c|}{\textbf{F}} & \multicolumn{1}{c|}{\textbf{$\eta^2$}} & \multicolumn{1}{c|}{\textbf{p}} \\ \hline\hline
		\textit{Profiling Method} & 51.79 & .42 & \textbf{.00} \\ \hline
    \textit{Decay Function} & 0.33 & .00 & .57 \\  \hline
    \textit{Document Content} & 5.16 & .04 & \textbf{.02} \\ \hline
    \textit{Profiling Method} $\times$ \textit{Decay Function} & 1.66 & .01 & .20 \\ \hline
    \textit{Profiling Method} $\times$ \textit{Document Content} & 4.76 & .02 & \textbf{.01} \\ \hline
    \textit{Decay Function} $\times$ \textit{Document Content} & 0.02 & .00 & .90 \\ \hline
		\breakcell{\textit{Profiling Method} $\times$ \textit{Decay Function} $\times$ \\  \textit{Document Content}} & 3.19 & .03 & \textbf{.04} \\ \hline
\end{tabular}
\label{tb:three-way-anova-map}
\end{table}
\paragraph*{\textbf{The factor \textit{Profiling Method}}}
Tables~\ref{tb:post-i-map}(a), (b) and (c) show the MAPs with respect to each profiling method, the post-hoc analysis for the factor \textit{Profiling Method}, and the effect size, respectively. 
Table~\ref{tb:post-i-map}(a) presents the means and standard deviations of the three profiling methods. 
Table~\ref{tb:post-i-map}(b) shows p-values of each pair. 
Since Table~\ref{tb:three-way-anova-map} shows that the factor \textit{Profiling Method} has the largest effect size, we further compute the effect size using Cohen's d for each pair shown in Table~\ref{tb:post-i-map}(c). 
The result shows that CF-IDF and HCF-IDF are superior to LDA. 
In contrast, there is no significant difference between CF-IDF and HCF-IDF, although MAP of HCF-IDF is slightly higher than CF-IDF.
\begin{table}[!ht]
\centering
\small
\caption{MAPs, Post-hoc analysis for the factor \textit{Profiling Method} using Shaffer's MSRB procedure, and effect size.}
\begin{tabular}{|c|c|c|}
		\multicolumn{3}{c}{\textbf{a) MAPs}} \\  \hline
    \multicolumn{1}{|c|}{\textbf{Choice}} & \multicolumn{1}{c|}{\textbf{M}} & \multicolumn{1}{c|}{\textbf{SD}} \\  \hline\hline
    HCF-IDF & .65 & .33 \\  \hline
		CF-IDF & .62 & .35 \\  \hline
	  LDA & .43 & .38 \\  \hline
		\multicolumn{3}{c}{\textbf{b) Post-hoc analysis p-values}} \\  \hline
     & HCF-IDF & LDA \\  \hline \hline
	  CF-IDF & .15 & \textbf{.00}  \\  \hline
    HCF-IDF & \cellcolor[HTML]{9B9B9B} & \textbf{.00} \\  \hline
		\multicolumn{3}{c}{\textbf{c) Effect size using Cohen's d}} \\  \hline
     & HCF-IDF & LDA \\  \hline \hline
	  CF-IDF & .09 & .52 \\  \hline
    HCF-IDF & \cellcolor[HTML]{9B9B9B} & .62 \\  \hline
\end{tabular}
\label{tb:post-i-map}
\end{table}
\paragraph*{\textbf{The factor \textit{Document Content}}}
Table~\ref{tb:post-iii-map} shows the post-hoc analysis for the factor \textit{Document Content}. 
It indicates that the recommender system works better when All (i\,.e., full-texts and titles) is taken into consideration for computing recommendations. 
\begin{table}[!ht]
\centering
\small
\caption{MAPs and Post-hoc analysis for the factor \textit{Document Content} using Shaffer's MSRB procedure.}
\begin{tabular}{|c|c|c|}
		\multicolumn{3}{c}{\textbf{a) MAPs}} \\  \hline
    \multicolumn{1}{|c|}{\textbf{Choice}} & \multicolumn{1}{c|}{\textbf{M}} & \multicolumn{1}{c|}{\textbf{SD}} \\  \hline\hline
		All & .59 & .38 \\  \hline
    Title & .55 & .35 \\  \hline
		\multicolumn{3}{c}{\textbf{b) Post-hoc analysis p-values}} \\  \hline
     & \multicolumn{2}{c|}{Title} \\  \hline \hline
	  All & \multicolumn{2}{c|}{\textbf{.02}}  \\  \hline
\end{tabular}
\label{tb:post-iii-map}
\end{table}
\paragraph*{\textbf{The factor \textit{Profiling Method} $\times$ \textit{Document Content}}}
Table~\ref{tb:anova-i-iii-map} shows the results of ANOVA regarding the factor \textit{Profiling Method} when a choice of the factor \textit{Document Content} is fixed and vice versa. 
We observe there are significant differences when a choice of the factor \textit{Document Content} is fixed and CF-IDF is employed. 
Mendoza's test found a violation of sphericity in the factor \textit{Profiling Method} when All is taken ($\chi^2(2) = 31.35$, $p=.00$). 
Thus, we run a one-way repeated-measure ANOVA with Greenhouse-Geisser correction of $\eta = .81$ for the second row in Table~\ref{tb:anova-i-iii-map}. 
Subsequently, we conduct the post-hoc analyses for each factor in which shows a significant difference. 
Table~\ref{tb:post-i-t-map} presents the post-hoc analysis when Title is employed. 
We see that HCF-IDF outperforms others with significant differences. 
Table~\ref{tb:post-i-f-map} shows the post-hoc analysis when All is chosen for the factor \textit{Document Content}. 
Different from the result shown in Table~\ref{tb:post-i-t-map}, CF-IDF performs slightly better than HCF-IDF, although there is no significant difference between them. 
Both CF-IDF and HCF-IDF demonstrate better recommendation performance than LDA. 
Table~\ref{tb:post-iii-c-map} shows the post-hoc analysis of the factor \textit{Document Content} when CF-IDF is employed. 
It indicates that the strategies with CF-IDF and All significantly outperforms those with CF-IDF and Title. 
\begin{table}[!ht]
\centering
\small
\caption{ANOVA for \textit{Profiling Method} $\times$ \textit{Document Content} interaction}
\begin{tabular}{|l|r|r|r|} \hline
    \multicolumn{1}{|c|}{\textbf{Factor}} & \multicolumn{1}{c|}{\textbf{F}} & \multicolumn{1}{c|}{\textbf{$\eta^2$}} & \multicolumn{1}{c|}{\textbf{p}} \\  \hline\hline
		\textit{Profiling Method} at Title & 23.99 & .20 & \textbf{.00} \\  \hline
    \textit{Profiling Method} at All & 36.35 & .30 & \textbf{.00} \\  \hline
    \textit{Document Content} at CF-IDF & 14.69 & .12 & \textbf{.00} \\  \hline
    \textit{Document Content} at HCF-IDF & 0.00 & .00 & .95 \\  \hline
    \textit{Document Content} at LDA & 0.01 & .00 & .93 \\  \hline
\end{tabular}
\label{tb:anova-i-iii-map}
\end{table}
\begin{table}[!ht]
\centering
\small
\caption{MAPs and Post-hoc analysis for the factor \textit{Profiling Method} at Title using Shaffer's MSRB procedure.}
\begin{tabular}{|c|c|c|}
		\multicolumn{3}{c}{\textbf{a) MAPs}} \\  \hline
    \multicolumn{1}{|c|}{\textbf{Choice}} & \multicolumn{1}{c|}{\textbf{M}} & \multicolumn{1}{c|}{\textbf{SD}} \\  \hline\hline
    HCF-IDF & .65 & .33 \\  \hline
		CF-IDF & .56 & .35 \\  \hline
	  LDA & .43 & .35 \\  \hline
		\multicolumn{3}{c}{\textbf{b) Post-hoc analysis p-values}} \\  \hline
        & HCF-IDF & LDA \\  \hline \hline
	  CF-IDF & \textbf{.01} & \textbf{.00}  \\  \hline
    HCF-IDF & \cellcolor[HTML]{9B9B9B} & \textbf{.00} \\  \hline
\end{tabular}
\label{tb:post-i-t-map}
\end{table}
\begin{table}[!ht]
\centering
\small
\caption{MAPs and Post-hoc analysis for the factor \textit{Profiling Method} at All using Shaffer's MSRB procedure.}
\begin{tabular}{|c|c|c|}
		\multicolumn{3}{c}{\textbf{a) MAPs}} \\  \hline
    \multicolumn{1}{|c|}{\textbf{Choice}} & \multicolumn{1}{c|}{\textbf{M}} & \multicolumn{1}{c|}{\textbf{SD}} \\  \hline\hline
		CF-IDF & .68 & .34 \\  \hline
    HCF-IDF & .65 & .34 \\  \hline
	  LDA & .44 & .41 \\  \hline
		\multicolumn{3}{c}{\textbf{b) Post-hoc analysis p-values}} \\  \hline
        & HCF-IDF & LDA \\  \hline \hline
	  CF-IDF & .21 & \textbf{.00}  \\  \hline
    HCF-IDF & \cellcolor[HTML]{9B9B9B} & \textbf{.00} \\  \hline
\end{tabular}
\label{tb:post-i-f-map}
\end{table}
\begin{table}[!ht]
\centering
\small
\caption{MAPs and Post-hoc analysis for the factor \textit{Document Content} at CF-IDF using Shaffer's MSRB procedure.}
\begin{tabular}{|c|c|c|}
		\multicolumn{3}{c}{\textbf{a) MAPs}} \\  \hline
    \multicolumn{1}{|c|}{\textbf{Choice}} & \multicolumn{1}{c|}{\textbf{M}} & \multicolumn{1}{c|}{\textbf{SD}} \\  \hline\hline
		All & .68 & .34 \\  \hline
    Title & .56 & .35 \\  \hline
		\multicolumn{3}{c}{\textbf{b) Post-hoc analysis p-values}} \\  \hline
        & \multicolumn{2}{c|}{All} \\  \hline \hline
	  Title & \multicolumn{2}{c|}{\textbf{.00}} \\  \hline
\end{tabular}
\label{tb:post-iii-c-map}
\end{table}
\section{Precision}
In this section, we evaluate the recommendation performance by computing Precision, especially Precision@5 (P@5). 
Precision is computed as described in Equation~\ref{eq:precision}. 
\begin{equation}
Precision@k = \frac{1}{k} \sum_{i = 1}^{k}{rel(i)}, 
\label{eq:precision}
\end{equation}
where $rel(k)$ returns $1$ if the item ranked at $i$ is relevant and $0$ if irrelevant. 
In this paper, we set $k = 5$, since five items are recommended by each strategy in the experiment. 
Using Precision@5, we first compare the twelve different strategies. 
Subsequently, we investigate the difference among the different experiment factors. 
\subsection{Best performing strategy}
Table~\ref{tb:precision} shows Precision@5 of each strategy. 
For the statistical analyses, we first applied Mauchly's test and found a violation of sphericity in the strategies ($\chi^2(65) = 421.32$, $p = .00$). 
Subsequently, we run a one-way repeated-measure ANOVA with a Greenhouse-Geisser correction of $\epsilon = .60$. 
It reveals a significant difference of the strategies' MAPs ($F(6.62, 808.00) = 21.85$, $p = .00$). 
To assess the statistical significance of pair-wise differences between the twelve strategies, a post-hoc analysis is performed using Shaffer's MSRB procedure~\cite{shaffer:1986}. 
The result of the post-hoc analysis is presented in Table~\ref{tb:pairwise-precision}. 
The vertical and horizontal dimensions of the Table~\ref{tb:pairwise-precision} show the eleven-by-eleven comparison of the twelve strategies. 
As one can see, we observe various significant differences between strategies (marked in bold font).
\begin{table}[!ht]
\centering
\small
\caption{Precision@5 (P@5) of the strategies in decreasing order.}
\begin{tabular}{|c|l|l|l||c|} \hline
		\multirow{2}{*}{} & \multicolumn{3}{c||}{\textbf{Strategy}} & \multicolumn{1}{c|}{\breakcell{\textbf{P@5}}} \\  \cline{2-5}
		& \multicolumn{1}{c|}{\breakcell{\textbf{Profiling}\\ \textbf{Method}}} & \multicolumn{1}{c|}{\breakcell{\textbf{Decay}\\ \textbf{Function}}} & \multicolumn{1}{c||}{\breakcell{\textbf{Con-} \\ \textbf{tent}}} & \multicolumn{1}{c|}{\textbf{M (SD)}} \\  \hline\hline
		1. & CF-IDF & Sliding window & All & .59 (.33) \\  \hline
		2. & HCF-IDF & Sliding window & All & .56 (.33) \\  \hline
		3. & HCF-IDF & Sliding window & Title & .55 (.33) \\  \hline
		4. & HCF-IDF & Exponential decay & Title & .52 (.30) \\  \hline
		5. & CF-IDF & Exponential decay & All & .50 (.32) \\  \hline
		6. & HCF-IDF & Exponential decay & All & .48 (.30) \\  \hline
		7. & CF-IDF & Exponential decay & Title & .40 (.29) \\  \hline
		8. & CF-IDF & Sliding window & Title & .39 (.27) \\  \hline
		9. & LDA & Exponential decay & Title & .37 (.31) \\  \hline
		10. & LDA & Sliding window & Title & .34 (.31) \\  \hline
		11. & LDA & Exponential decay & All & .31 (.30) \\  \hline
		12. & LDA & Sliding window & All & .27 (.33) \\  \hline
\end{tabular}
\label{tb:precision}
\end{table}

\begin{table*}[!ht]
\centering
\caption{Post-hoc analysis of Precision@5 (P@5) with pairwise p-values over the strategies using Shaffer's MSRB procedure. The p-values are marked in bold font if $p < .05$, which indicates a significant difference between the two strategies. Strategies are sorted by Precision@5 as shown in Table~\ref{tb:precision}.}
\small
\begin{tabular}{|c|l|l|l||c|c|c|c|c|c|c|c|c|c|c|} \cline{5-15}
		\multicolumn{4}{l|}{\multirow{4}{*}{}} & \rot{All} & \rot{Title} & \rot{Title} & \rot{All} & \rot{All} & \rot{Title} & \rot{Title} & \rot{Title} & \rot{Title} & \rot{All} & \rot{All} \\  \cline{5-15}
		\multicolumn{4}{c|}{} & \rot{\breakcell{Sliding\\window}} & \rot{\breakcell{Sliding\\window}} & \rot{\breakcell{Exponen-\\tial decay}} & \rot{\breakcell{Exponen-\\tial decay}} & \rot{\breakcell{Exponen-\\tial decay}} & \rot{\breakcell{Exponen-\\tial decay}} & \rot{\breakcell{Sliding\\window}} & \rot{\breakcell{Exponen-\\tial decay}} & \rot{\breakcell{Sliding\\window}} & \rot{\breakcell{Exponen-\\tial decay}} & \rot{\breakcell{Sliding\\window}} \\  \cline{5-15}
		\multicolumn{4}{c|}{} & \rot{HCF-IDF} & \rot{HCF-IDF} & \rot{HCF-IDF} & \rot{CF-IDF} & \rot{HCF-IDF} & \rot{CF-IDF} & \rot{CF-IDF} & \rot{LDA} & \rot{LDA} & \rot{LDA} & \rot{LDA} \\  \cline{5-15}
		\multicolumn{4}{c|}{} & 2. & 3. & 4. & 5. & 6. & 7. & 8. & 9. & 10. & 11. & 12. \\  \hhline{----*{10}{|=}|=|}
		1. & CF-IDF & Sliding window & All &  .99 & .99 & .60 & .09 & \textbf{.05} & \textbf{.00} & \textbf{.00} & \textbf{.00} & \textbf{.00} & \textbf{.00} & \textbf{.00} \\  \hline
		2. & HCF-IDF & Sliding window & All & \cellcolor[HTML]{9B9B9B} & .99 & .99 & .99 & .50 & \textbf{.00} & \textbf{.00} & \textbf{.00} & \textbf{.00} & \textbf{.00} & \textbf{.00} \\  \hline
		3. & HCF-IDF & Sliding window & Title &  \cellcolor[HTML]{9B9B9B} & \cellcolor[HTML]{9B9B9B} & .99 & .99 & .99 & \textbf{.00} & \textbf{.00} & \textbf{.00} & \textbf{.00} & \textbf{.00} & \textbf{.00} \\  \hline
		4. & HCF-IDF & Exponential decay & Title & \cellcolor[HTML]{9B9B9B} & \cellcolor[HTML]{9B9B9B} & \cellcolor[HTML]{9B9B9B} & .99 & .99 & \textbf{.00} & \textbf{.00} & \textbf{.00} & \textbf{.00} & \textbf{.00} & \textbf{.00} \\  \hline
		5. & CF-IDF & Exponential decay & All & \cellcolor[HTML]{9B9B9B} & \cellcolor[HTML]{9B9B9B} & \cellcolor[HTML]{9B9B9B} & \cellcolor[HTML]{9B9B9B} & .99 & \textbf{.03} & \textbf{.01} & \textbf{.02} & \textbf{.00} & \textbf{.00} & \textbf{.00} \\  \hline
		6. & HCF-IDF & Exponential decay & All & \cellcolor[HTML]{9B9B9B} & \cellcolor[HTML]{9B9B9B} & \cellcolor[HTML]{9B9B9B} & \cellcolor[HTML]{9B9B9B} & \cellcolor[HTML]{9B9B9B} & .09 & \textbf{.03} & \textbf{.03} & \textbf{.00} & \textbf{.00} & \textbf{.00} \\  \hline
    7. & CF-IDF & Exponential decay & Title &  \cellcolor[HTML]{9B9B9B} & \cellcolor[HTML]{9B9B9B} & \cellcolor[HTML]{9B9B9B} & \cellcolor[HTML]{9B9B9B} & \cellcolor[HTML]{9B9B9B} & \cellcolor[HTML]{9B9B9B} & .99 & .99 & .99 & .26 & \textbf{.01} \\  \hline
    8. & CF-IDF & Sliding window & Title & \cellcolor[HTML]{9B9B9B} & \cellcolor[HTML]{9B9B9B} & \cellcolor[HTML]{9B9B9B} & \cellcolor[HTML]{9B9B9B} & \cellcolor[HTML]{9B9B9B} & \cellcolor[HTML]{9B9B9B} & \cellcolor[HTML]{9B9B9B} & .99 & .99 & .34 & \textbf{.02} \\  \hline
    9. & LDA & Exponential decay & Title & \cellcolor[HTML]{9B9B9B} & \cellcolor[HTML]{9B9B9B} & \cellcolor[HTML]{9B9B9B} & \cellcolor[HTML]{9B9B9B} & \cellcolor[HTML]{9B9B9B} & \cellcolor[HTML]{9B9B9B} & \cellcolor[HTML]{9B9B9B} & \cellcolor[HTML]{9B9B9B} & .99 & .99 & .09 \\  \hline
		10. & LDA & Sliding window & Title & \cellcolor[HTML]{9B9B9B} & \cellcolor[HTML]{9B9B9B} & \cellcolor[HTML]{9B9B9B} & \cellcolor[HTML]{9B9B9B} & \cellcolor[HTML]{9B9B9B} & \cellcolor[HTML]{9B9B9B} & \cellcolor[HTML]{9B9B9B} & \cellcolor[HTML]{9B9B9B} & \cellcolor[HTML]{9B9B9B} & .99 & .82 \\  \hline
		11. & LDA & Exponential decay & All & \cellcolor[HTML]{9B9B9B} & \cellcolor[HTML]{9B9B9B} & \cellcolor[HTML]{9B9B9B} & \cellcolor[HTML]{9B9B9B} & \cellcolor[HTML]{9B9B9B} & \cellcolor[HTML]{9B9B9B} & \cellcolor[HTML]{9B9B9B} & \cellcolor[HTML]{9B9B9B} & \cellcolor[HTML]{9B9B9B} & \cellcolor[HTML]{9B9B9B} & .99 \\  \hline
\end{tabular}
\label{tb:pairwise-precision}
\end{table*}
\subsection{Difference in experiment factors}
Subsequently, we analyze the results of Precision@5 with respect to each factor. 
First, we apply Mendoza's test~\cite{mendoza:1980} which showed violations of sphericity against the factors \textit{Profiling Method} ($\chi^2(2) = 13.92$, $p = .00$), \textit{Profiling Method} $\times$ \textit{Decay Function} ($\chi^2(2) = 19.64$, $p = .00$), and \textit{Profiling Method} $\times$ \textit{Document Content} ($\chi^2(2) = 7.23$, $p = .03$). 
Thus, we run three-way repeated-measure ANOVA with a Greenhouse-Geisser correction of $\epsilon = .90$ for the factor \textit{Profiling Method}, $\epsilon = .87$ for the factor \textit{Profiling Method} $\times$ \textit{Decay Function}, and $\epsilon = .95$ for the factor \textit{Profiling Method} $\times$ \textit{Document Content}. 
Table~\ref{tb:three-way-anova-precision} shows the result of an ANOVA with F-ratio, $\eta^2$ and p-value. 
$\eta^2$ indicates the effect size of each factor. 
The effect size is small when $\eta^2 > .02$, medium when $\eta^2 > .13$, and large when $\eta^2 > .26$. 
For all factors that make significant difference, we conduct a post-hoc analysis using Shaffer's MSRB Procedure. 
\begin{table}[!ht]
\centering
\small
\caption{Three-way repeated-measure ANOVA with Greenhouse-Geisser correction with F-ratio, $\eta^2$ and p-value for Precision@5.}
\begin{tabular}{|l|r|r|r|} \hline
    \multicolumn{1}{|c|}{\textbf{Factor}} & \multicolumn{1}{c|}{\textbf{F}} & \multicolumn{1}{c|}{\textbf{$\eta^2$}} & \multicolumn{1}{c|}{\textbf{p}} \\  \hline\hline
		\textit{Profiling Method} & 54.24 & .42 & \textbf{.00} \\  \hline
    \textit{Decay Function} & 1.75 & .00 & .19 \\  \hline
    \textit{Document Content} & 3.23 & .04 & .08 \\  \hline
    \textit{Profiling Method} $\times$ \textit{Decay Function} & 6.32 & .01 & \textbf{.00} \\  \hline
    \textit{Profiling Method} $\times$ \textit{Document Content} & 20.53 & .02 & \textbf{.00} \\  \hline
    \textit{Decay Function} $\times$ \textit{Document Content} & 7.13 & .00 & \textbf{.01} \\  \hline
		\breakcell{\textit{Profiling Method} $\times$ \textit{Decay Function} $\times$ \\ \textit{Document Content}} & 2.61 & .03 & .07 \\  \hline
\end{tabular}
\label{tb:three-way-anova-precision}
\end{table}
\paragraph*{\textbf{The factor \textit{Profiling Method}}}
Tables~\ref{tb:post-i-precision}(a), (b) and (c) show the Precision@5, the post-hoc analysis for the factor \textit{Profiling Method}, and the effect size, respectively. 
Table~\ref{tb:post-i-precision}(a) presents the means and standard deviations of the three profiling methods. 
Table~\ref{tb:post-i-precision}(b) shows p-values of each pair. 
Since Table~\ref{tb:three-way-anova-precision} shows that this factor has the largest effect size, we further compute the effect size using Cohen's d for each pair shown in Table~\ref{tb:post-i-map}(c). 
There are significant differences between all pairs of the three profiling methods and among the three profiling methods HCF-IDF performs best significantly. 
\begin{table}[!ht]
\centering
\small
\caption{Precision@5, Post-hoc analysis for the factor \textit{Profiling Method} using Shaffer's MSRB procedure, and effect size.}
\begin{tabular}{|c|c|c|}
		\multicolumn{3}{c}{\textbf{a) Precision@5}} \\  \hline
    \multicolumn{1}{|c|}{\textbf{Choice}} & \multicolumn{1}{c|}{\textbf{M}} & \multicolumn{1}{c|}{\textbf{SD}} \\  \hline\hline
    HCF-IDF & .53 & .31 \\  \hline
		CF-IDF & .47 & .31 \\  \hline
	  LDA & .32 & .31 \\  \hline
		\multicolumn{3}{c}{\textbf{b) Post-hoc analysis p-values}} \\  \hline
     & HCF-IDF & LDA \\  \hline \hline
	  CF-IDF & \textbf{.00} & \textbf{.00}  \\  \hline
    HCF-IDF & \cellcolor[HTML]{9B9B9B} & \textbf{.00} \\  \hline
		\multicolumn{3}{c}{\textbf{c) Effect size using Cohen's d}} \\  \hline
     & HCF-IDF & LDA \\  \hline \hline
	  CF-IDF & .09 & .52 \\  \hline
    HCF-IDF & \cellcolor[HTML]{9B9B9B} & .62 \\  \hline
\end{tabular}
\label{tb:post-i-precision}
\end{table}
\paragraph*{\textbf{The factor \textit{Profiling Method} $\times$ \textit{Decay Function}}}
Table~\ref{tb:anova-i-ii-precision} shows the results of ANOVA regarding the \textit{profiling method} when a choice of the \textit{Decay Function} is fixed and vice versa. 
There are significant differences when the choice of the factor \textit{Decay Function} is fixed. 
In both decay functions, all pairs of the three profiling methods show significant differences. 
Specifically, HCF-IDF performs best, followed by CF-IDF and LDA. 
When CF-IDF is employed, Sliding window makes significantly better recommendations than Exponential decay ($F(1,122) = 5.44$, $p = .02$). 
In contrast, when LDA is employed, Exponential decay performs significantly better than Sliding window ($F(1,122) = 6.75$, $p = .01$).
The factor \textit{Decay Function} does not make difference on the recommendation performance when HCF-IDF is employed. 
\begin{table}[!ht]
\centering
\small
\caption{ANOVA for \textit{Profiling Method} $\times$ \textit{Decay Function} interaction}
\begin{tabular}{|l|r|r|r|} \hline
    \multicolumn{1}{|c|}{\textbf{Factor}} & \multicolumn{1}{c|}{\textbf{F}} & \multicolumn{1}{c|}{\textbf{$\eta^2$}} & \multicolumn{1}{c|}{\textbf{p}} \\  \hline\hline
		\textit{Profiling Method} at Sliding window & 52.98 & .20 & \textbf{.00} \\  \hline
    \textit{Profiling Method} at Exponential decay & 22.52 & .30 & \textbf{.00} \\  \hline
    \textit{Decay Function} at CF-IDF & 5.44 & .12 & \textbf{.02} \\  \hline
    \textit{Decay Function} at HCF-IDF & 3.25 & .00 & .07 \\  \hline
    \textit{Decay Function} at LDA & 6.75 & .00 & \textbf{.01} \\  \hline
\end{tabular}
\label{tb:anova-i-ii-precision}
\end{table}
\paragraph*{\textbf{The factor \textit{Profiling Method} $\times$ \textit{Document Content}}}
Table~\ref{tb:anova-i-iii-precision} shows the results of ANOVA regarding the factor \textit{Profiling Method} when a choice of the \textit{Document Content} is fixed and vice versa. 
When the choice of the \textit{Document Content} is Title, HCF-IDF performs best and significantly better than both CF-IDF and LDA. 
There is no significant difference between CF-IDF and LDA. 
When the choice of the \textit{Document Content} is All, HCF-IDF performs best. 
But, there is no significant difference between CF-IDF and HCF-IDF and both profiling methods are significantly superior to LDA. 
When CF-IDF is employed, All is the better choice than Title. 
In contrast, Title performs better than All, when LDA is employed. 
\begin{table}[!ht]
\centering
\small
\caption{ANOVA for \textit{Profiling Method} $\times$ \textit{Document Content} (\textit{Document Content}) interaction}
\begin{tabular}{|l|r|r|r|} \hline
    \multicolumn{1}{|c|}{\textbf{Factor}} & \multicolumn{1}{c|}{\textbf{F}} & \multicolumn{1}{c|}{\textbf{$\eta^2$}} & \multicolumn{1}{c|}{\textbf{p}} \\  \hline\hline
		\textit{Profiling Method} at Title & 23.37 & .20 & \textbf{.00} \\  \hline
    \textit{Profiling Method} at All & 56.54 & .30 & \textbf{.00} \\  \hline
    \textit{Document Content} at CF-IDF & 33.39 & .12 & \textbf{.00} \\  \hline
    \textit{Document Content} at HCF-IDF & 0.44 & .00 & .51 \\  \hline
    \textit{Document Content} at LDA & 4.68 & .00 & \textbf{.03} \\  \hline
\end{tabular}
\label{tb:anova-i-iii-precision}
\end{table}
\paragraph*{\textbf{The factor \textit{Decay Function} $\times$ \textit{Document Content}}}
Table~\ref{tb:anova-ii-iii-precision} shows the results of ANOVA regarding the factor \textit{Decay Function} when a choice of the factor \textit{Document Content} is fixed and vice versa. 
When All is choosen for the factor \textit{Document Content}, Sliding window is the better decay function. 
When Sliding window is employed in the strategies, the strategies with All is significantly better than those with Title.
\begin{table}[!ht]
\centering
\small
\caption{ANOVA for \textit{Decay Function} $\times$ \textit{Document Content} interaction}
\begin{tabular}{|l|r|r|r|} \hline
    \multicolumn{1}{|c|}{\textbf{Factor}} & \multicolumn{1}{c|}{\textbf{F}} & \multicolumn{1}{c|}{\textbf{$\eta^2$}} & \multicolumn{1}{c|}{\textbf{p}} \\  \hline\hline
		\textit{Decay Function} at Title & 0.08 & .20 & .78 \\  \hline
    \textit{Decay Function} at All & 4.99 & .30 & \textbf{.03} \\  \hline
    \textit{Document Content} at Sliding window & 8.74 & .12 & \textbf{.00} \\  \hline
    \textit{Document Content} at Exponential decay & 0.00 & .00 & .97 \\  \hline
\end{tabular}
\label{tb:anova-ii-iii-precision}
\end{table}
\section{Mean Reciprocal Rank}
In this section, we evaluate the recommendation performance by computing Mean Reciprocal Rank (MRR). 
Reciprocal Rank is defined as Equation~\ref{eq:rr}. 
\begin{equation}
RR = \frac{1}{rank_{first}}, 
\label{eq:rr}
\end{equation}
where $rank_{first}$ indicates the rank position of the first item which is evaluated as interesting. 
Mean Reciprocal Rank (MRR) is the mean of the Reciprocal Rank scores for each participant. 
If there is no relevant items in the recommendation list, RR outputs $0$. 
Using MRR, we first compare the twelve different strategies. 
Subsequently, we investigate the difference among the different experiment factors. 
\subsection{Best Performing Strategy}
Table~\ref{tb:mrr} shows the Mean Reciprocal Ranks (MRR) of each strategies. 
The order of the strategies are different from rankscores shown in~Table \ref{tb:single-all}. 
For the statistical analyses, we first applied Mauchly's test and found a violation of sphericity in the strategies ($\chi^2(65) = 308.70$, $p = .00$). 
Subsequently, we ran a one-way repeated-measure ANOVA with a Greenhouse-Geisser correction of $\epsilon = .67$. 
It revealed a significant difference of the strategies' MRRs ($F(0.18, 2.53) = 14.40$, $p = .00$). 
To assess the statistical significance of pair-wise differences between the twelve strategies, a post-hoc analysis was performed using Shaffer's MSRB procedure~\cite{shaffer:1986}. 
The result of the post-hoc analysis is presented in Table~\ref{tb:pairwise-mrr}. The vertical and horizontal dimensions of the Table~\ref{tb:pairwise-mrr} show the eleven-by-eleven comparison of the twelve strategies. 
As one can see, we observe various significant differences between strategies (marked in bold font).
\begin{table}[!ht]
\centering
\small
\caption{Mean Reciprocal Rank (MRR) of the strategies in decreasing order.}
\begin{tabular}{|c|l|l|l||c|} \hline
\multirow{2}{*}{} & \multicolumn{3}{c||}{\textbf{Strategy}} & \multicolumn{1}{c|}{\breakcell{\textbf{MRR}}} \\  \cline{2-5}
		& \multicolumn{1}{c|}{\breakcell{\textbf{Profiling}\\ \textbf{Method}}} & \multicolumn{1}{c|}{\breakcell{\textbf{Decay}\\ \textbf{Function}}} & \multicolumn{1}{c||}{\breakcell{\textbf{Con-} \\ \textbf{tent}}} & \multicolumn{1}{c|}{\textbf{M (SD)}} \\  \hline\hline
		1 & CF-IDF & Sliding window & All & .73 (.35) \\  \hline
		2 & CF-IDF & Exponential decay & All & .69 (.39) \\  \hline
		3 & HCF-IDF & Exponential decay & All & .68 (.37) \\  \hline
		4 & HCF-IDF & Exponential decay & Title & .68 (.37) \\  \hline
		5 & HCF-IDF & Sliding window & Title & .67 (.38) \\  \hline
		6 & HCF-IDF & Sliding window & All & .67 (.37) \\  \hline
		7 & CF-IDF & Exponential decay & Title & .61 (.39) \\  \hline
		8 & CF-IDF & Sliding window & Title & .59 (.39) \\  \hline
		9 & LDA & Exponential decay & All & .50 (.43) \\  \hline
		10 & LDA & Exponential decay & Title & .43 (.37) \\  \hline
		11 & LDA & Sliding window & Title & .42 (.38) \\  \hline
		12 & LDA & Sliding window & All & .41 (.44) \\  \hline
\end{tabular}
\label{tb:mrr}
\end{table}
\begin{table*}[!ht]
\small
\centering
\caption{Post-hoc analysis of Mean Reciprocal Rank (MRR) with pairwise p-values over the strategies using Shaffer's MSRB procedure. The p-values are marked in bold font if $p < .05$, which indicates a significant difference between the two strategies. Strategies are sorted by MRR as shown in Table~\ref{tb:mrr}.}
\small
\begin{tabular}{|c|l|l|l||c|c|c|c|c|c|c|c|c|c|c|c|} \cline{5-15}
		\multicolumn{4}{l|}{\multirow{4}{*}{}} & \rot{All} & \rot{All} & \rot{Title} & \rot{Title} & \rot{All} & \rot{Title} & \rot{Title} & \rot{All} & \rot{Title} & \rot{Title} & \rot{All} \\  \cline{5-15}
		\multicolumn{4}{c|}{} & \rot{\breakcell{Exponen-\\tial decay}} & \rot{\breakcell{Exponen-\\tial decay}} & \rot{\breakcell{Exponen-\\tial decay}} & \rot{\breakcell{Sliding\\window}} & \rot{\breakcell{Sliding\\window}} & \rot{\breakcell{Exponen-\\tial decay}} & \rot{\breakcell{Sliding\\window}} & \rot{\breakcell{Exponen-\\tial decay}} & \rot{\breakcell{Exponen-\\tial decay}} & \rot{\breakcell{Sliding\\window}} & \rot{\breakcell{Sliding\\window}} \\  \cline{5-15}
		\multicolumn{4}{c|}{} & \rot{CF-IDF} & \rot{HCF-IDF} & \rot{HCF-IDF} & \rot{HCF-IDF} & \rot{HCF-IDF} & \rot{CF-IDF} & \rot{CF-IDF} & \rot{LDA} & \rot{LDA} & \rot{LDA} & \rot{LDA} \\  \cline{5-15}
		\multicolumn{4}{c|}{} & 2. & 3. & 4. & 5. & 6. & 7. & 8. & 9. & 10. & 11. & 12. \\  \hhline{----*{10}{|=}|=|}
    1. & CF-IDF & Sliding window & All & .99 & .99 & .99 & .99 & .99 & .12 & \textbf{.03} & \textbf{.00} & \textbf{.00} & \textbf{.00} & \textbf{.00} \\  \hline
    2. & CF-IDF & Exponential decay & All & \cellcolor[HTML]{9B9B9B} & .99 & .99 & .99 & .99 & .99 & .99 & \textbf{.01} & \textbf{.00} & \textbf{.00} & \textbf{.00} \\  \hline
		3. & HCF-IDF & Exponential decay & All & \cellcolor[HTML]{9B9B9B} & \cellcolor[HTML]{9B9B9B} & .99 & .99 & .99 & .99 & .99 & \textbf{.01} & \textbf{.00} & \textbf{.00} & \textbf{.01} \\  \hline
		4. & HCF-IDF & Exponential decay & Title & \cellcolor[HTML]{9B9B9B} & \cellcolor[HTML]{9B9B9B} & \cellcolor[HTML]{9B9B9B} & .99 & .99 & .99 & .99 & \textbf{.01} & \textbf{.00} & \textbf{.00} & \textbf{.00} \\  \hline
    5. & HCF-IDF & Sliding window & Title & \cellcolor[HTML]{9B9B9B} & \cellcolor[HTML]{9B9B9B} & \cellcolor[HTML]{9B9B9B} & \cellcolor[HTML]{9B9B9B} & .99 & .99 & .99 & \textbf{.01} & \textbf{.00} & \textbf{.00} & \textbf{.00} \\  \hline
    6. & HCF-IDF & Sliding window & All & \cellcolor[HTML]{9B9B9B} & \cellcolor[HTML]{9B9B9B} & \cellcolor[HTML]{9B9B9B} & \cellcolor[HTML]{9B9B9B} & \cellcolor[HTML]{9B9B9B} & .99 & .99 & \textbf{.01} & \textbf{.00} & \textbf{.00} & \textbf{.01} \\  \hline
		7. & CF-IDF & Exponential decay & Title & \cellcolor[HTML]{9B9B9B} & \cellcolor[HTML]{9B9B9B} & \cellcolor[HTML]{9B9B9B} & \cellcolor[HTML]{9B9B9B} & \cellcolor[HTML]{9B9B9B} & \cellcolor[HTML]{9B9B9B} & .99 & .94 & \textbf{.00} & \textbf{.00} & \textbf{.00} \\  \hline
		8. & CF-IDF & Sliding window & Title & \cellcolor[HTML]{9B9B9B} & \cellcolor[HTML]{9B9B9B} & \cellcolor[HTML]{9B9B9B} & \cellcolor[HTML]{9B9B9B} & \cellcolor[HTML]{9B9B9B} & \cellcolor[HTML]{9B9B9B} & \cellcolor[HTML]{9B9B9B} & .99 & \textbf{.00} & \textbf{.00} & \textbf{.02}  \\  \hline
		9. & LDA & Exponential decay & All & \cellcolor[HTML]{9B9B9B} & \cellcolor[HTML]{9B9B9B} & \cellcolor[HTML]{9B9B9B} & \cellcolor[HTML]{9B9B9B} & \cellcolor[HTML]{9B9B9B} & \cellcolor[HTML]{9B9B9B} & \cellcolor[HTML]{9B9B9B} & \cellcolor[HTML]{9B9B9B} & .99 & .99 & .58 \\  \hline
		10. & LDA & Exponential decay & Title & \cellcolor[HTML]{9B9B9B} & \cellcolor[HTML]{9B9B9B} & \cellcolor[HTML]{9B9B9B} & \cellcolor[HTML]{9B9B9B} & \cellcolor[HTML]{9B9B9B} & \cellcolor[HTML]{9B9B9B} & \cellcolor[HTML]{9B9B9B} & \cellcolor[HTML]{9B9B9B} & \cellcolor[HTML]{9B9B9B} & .99 & .99 \\  \hline
		11. & LDA & Sliding window & Title & \cellcolor[HTML]{9B9B9B} & \cellcolor[HTML]{9B9B9B} & \cellcolor[HTML]{9B9B9B} & \cellcolor[HTML]{9B9B9B} & \cellcolor[HTML]{9B9B9B} & \cellcolor[HTML]{9B9B9B} & \cellcolor[HTML]{9B9B9B} & \cellcolor[HTML]{9B9B9B} & \cellcolor[HTML]{9B9B9B} & \cellcolor[HTML]{9B9B9B} & .99 \\  \hline
\end{tabular}
\label{tb:pairwise-mrr}
\end{table*}

\subsection{Difference in experiment factors}
Subsequently, we look into the results of MRRs with respect to each factor. 
First, we applied Mendoza's test~\cite{mendoza:1980} which showed violations of sphericity against the factors \textit{Profiling Method} $\times$ \textit{Decay Function} ($\chi^2(2) = 8.16$, $p = .02$), and \textit{Profiling Method} $\times$ \textit{Document Content} ($\chi^2(2) = 8.85$, $p = .01$). 
Thus, we ran three-way repeated-measure ANOVA with a Greenhouse-Geisser correction of $\epsilon = .94$ for \textit{Profiling Method} $\times$ \textit{Decay Function}, and $\epsilon = .93$ for \textit{profiling method} $\times$ \textit{document content}. 
Table~\ref{tb:three-way-anova-mrr} shows the results of an ANOVA with F-ratio, $\eta^2$ and p-value. 
The analysis revealed significant differences only in the two factors \textit{Profiling Method} and \textit{Document Content}. 
\begin{table}[!ht]
\centering
\small
\caption{Three-way repeated-measure ANOVA with Greenhouse-Geisser correction with F-ratio, $\eta^2$ and p-value for MRR.}
\begin{tabular}{|l|r|r|r|} \hline
    \multicolumn{1}{|c|}{\textbf{Factor}} & \multicolumn{1}{c|}{\textbf{F}} & \multicolumn{1}{c|}{\textbf{$\eta^2$}} & \multicolumn{1}{c|}{\textbf{p}} \\  \hline\hline
		\textit{Profiling Method} & 50.65 & .42 & \textbf{.00} \\  \hline
    \textit{Decay Function} & 0.56 & .00 & .45 \\  \hline
    \textit{Document Content} & 5.10 & .04 & \textbf{.03} \\  \hline
    \textit{Profiling Method} $\times$ \textit{Decay Function} & 1.28 & .01 & .28 \\  \hline
    \textit{Profiling Method} $\times$ \textit{Document Content} & 2.83 & .02 & .06 \\  \hline
    \textit{Decay Function} $\times$ \textit{Document Content} & 0.13 & .00 & .72 \\  \hline
		\breakcell{\textit{Profiling Method} $\times$ \textit{Decay Function} $\times$ \\ \textit{Document Content}} & 2.33 & .02 & .10 \\  \hline
\end{tabular}
\label{tb:three-way-anova-mrr}
\end{table}
\paragraph*{\textbf{The factor \textit{Profiling Method}}}
Tables~\ref{tb:post-i-mrr}(a), (b) and (c) show the MRRs, the post-hoc analysis for the factor \textit{Profiling Method}, and the effect size, respectively. 
Table~\ref{tb:post-i-mrr}(a) presents the means and standard deviations of the three profiling methods. 
Table~\ref{tb:post-i-mrr}(b) shows p-values of each pair. 
Since Table~\ref{tb:three-way-anova-mrr} shows that this factor has the largest effect size, we further compute the effect size using Cohen's d for each pair shown in Table~\ref{tb:post-i-mrr}(c). 
%
\begin{table}[!ht]
\centering
\small
\caption{MRRs, Post-hoc analysis for the factor \textit{Profiling Method} using Shaffer's MSRB procedure, and effect size.}
\begin{tabular}{|c|c|c|}
		\multicolumn{3}{c}{\textbf{a) MRRs}} \\  \hline
    \multicolumn{1}{|c|}{\textbf{Choice}} & \multicolumn{1}{c|}{\textbf{M}} & \multicolumn{1}{c|}{\textbf{SD}} \\  \hline\hline
    HCF-IDF & .68 & .38 \\  \hline
		CF-IDF & .66 & .37 \\  \hline
	  LDA & .44 & .41 \\  \hline
		\multicolumn{3}{c}{\textbf{b) Post-hoc analysis p-values}} \\  \hline
     & HCF-IDF & LDA \\  \hline \hline
	  CF-IDF & .34 & \textbf{.00}  \\  \hline
    HCF-IDF & \cellcolor[HTML]{9B9B9B} & \textbf{.00} \\  \hline
		\multicolumn{3}{c}{\textbf{c) Effect size using Cohen's d}} \\  \hline
     & HCF-IDF & LDA \\  \hline \hline
	  CF-IDF & .05 & .56 \\  \hline
    HCF-IDF & \cellcolor[HTML]{9B9B9B} & .61 \\  \hline
\end{tabular}
\label{tb:post-i-mrr}
\end{table}
\paragraph*{The factor \textit{Document Content}}
Table~\ref{tb:post-iii-mrr} shows the post-hoc analysis for the factor \textit{Document Content}. 
It indicates that generally the recommender system work better when full-texts are available. 
\begin{table}[!ht]
\centering
\small
\caption{MRRs and Post-hoc analysis for the factor \textit{Document Content} using Shaffer's MSRB procedure.}
\begin{tabular}{|c|c|c|}
		\multicolumn{3}{c}{\textbf{a) MRRs}} \\  \hline
    \multicolumn{1}{|c|}{\textbf{Choice}} & \multicolumn{1}{c|}{\textbf{M}} & \multicolumn{1}{c|}{\textbf{SD}} \\  \hline\hline
		All & .61 & .41 \\  \hline
    Title & .57 & .39 \\  \hline
		\multicolumn{3}{c}{\textbf{b) Post-hoc analysis p-values}} \\  \hline
     & \multicolumn{2}{c|}{All} \\  \hline \hline
	  Title & \multicolumn{2}{c|}{\textbf{.03}}  \\  \hline
\end{tabular}
\label{tb:post-iii-mrr}
\end{table}
\section{Normalized Discounted Cumulative Gain}
In this section, we evaluate the recommendation performance by Normalized Discounted Cumulative Gain (nDCG). 
Discounted Cumulative Gain (DCG) is calculated as Equation~\ref{eq:dcg}. 
\begin{equation}
DCG = \sum_{i = 1}^{k}{\frac{2^{rel(i)} - 1}{\log_{2}{i}}}, 
\label{eq:dcg}
\end{equation}
where $rel(k)$ returns $1$ if the item ranked at $i$ is relevant and $0$ if irrelevant. 
Similar to rankscore, the items ranked at higher positions have a larger influence on output score. 
First, we compare the twelve different strategies using the metric. 
Subsequently, we investigate the difference among the different experiment factors. 
\subsection{Best performing strategy}
Table~\ref{tb:ndcg} shows the Normalized Discounted Cumulative Gain (nDCG) of the twelve strategies. 
The order of the strategies is identical with rankscores shown in Table~\ref{tb:single-all}. 
For the statistical analyses, we first apply Mauchly's test and found a violation of sphericity in the strategies ($\chi^2(65) = 424.00$, $p = .00$). 
Subsequently, we run a one-way repeated-measure ANOVA with a Greenhouse-Geisser correction of $\epsilon = .61$. 
It reveals a significant difference of the strategies' nDCG ($F(6.69, 816.37) = 21.16$, $p = .00$). 
To assess the statistical significance of pair-wise differences between the twelve strategies, a post-hoc analysis is performed using Shaffer's MSRB procedure~\cite{shaffer:1986}. 
The result of the post-hoc analysis is presented in Table~\ref{tb:pairwise-ndcg}. 
The vertical and horizontal dimensions of the Table~\ref{tb:pairwise-ndcg} show the eleven-by-eleven comparison of the twelve strategies. 
As one can see, we observe various significant differences between strategies (marked in bold font).
\begin{table}[!ht]
\centering
\small
\caption{nDCGs of the strategies in decreasing order. M and SD denote mean and standard deviation, respectively.}
\begin{tabular}{|c|l|l|l||c|} \hline
		\multirow{2}{*}{} & \multicolumn{3}{c||}{\textbf{Strategy}} & \multicolumn{1}{c|}{\textbf{nDCG}} \\ \cline{2-5}
		& \multicolumn{1}{c|}{\breakcell{\textbf{Profiling}\\ \textbf{Method}}} & \multicolumn{1}{c|}{\breakcell{\textbf{Decay}\\ \textbf{Function}}} & \multicolumn{1}{c||}{\breakcell{\remove{\textbf{Doc-}\\ \textbf{ment} \\ }\textbf{Con-} \\ \textbf{tent}}} & \multicolumn{1}{c|}{\textbf{M (SD)}} \\  \hline\hline
		1. & CF-IDF & Sliding window & All & .59 (.33)\\  \hline
		2. & HCF-IDF & Sliding window & All & .56 (.34)\\  \hline
		3. & HCF-IDF & Sliding window & Title & .55 (.33)\\  \hline
		4. & HCF-IDF & Exponential decay & Title & .52 (.30)\\  \hline
		5. & CF-IDF & Exponential decay & All & .52 (.32)\\  \hline
		6. & HCF-IDF & Exponential decay & All & .50 (.30)\\  \hline
		7. & CF-IDF & Exponential decay & Title & .41 (.30)\\  \hline
		8. & CF-IDF & Sliding window & Title & .40 (.27)\\  \hline
		9. & LDA & Exponential decay & Title & .34 (.31)\\  \hline
		10. & LDA & Sliding window & Title & .32 (.31)\\  \hline
		11. & LDA & Exponential decay & All & .32 (.31)\\  \hline
		12. & LDA & Sliding window & All & .28 (.33)\\  \hline
\end{tabular}
\label{tb:ndcg}
\end{table}
%
\begin{table*}[!ht]
\centering
\caption{Post-hoc analysis of normalized Discounted Cumulative Gain (nDCG) with pairwise p-values over the twelve strategies using Shaffer's MSRB procedure. The p-values are marked in bold font if $p < .05$, which indicates a significant difference between the two strategies. Strategies are sorted by nDCG as shown in Table~\ref{tb:ndcg}.}
\small
\begin{tabular}{|c|l|l|l||c|c|c|c|c|c|c|c|c|c|c|} \cline{5-15}
		\multicolumn{4}{l|}{\multirow{4}{*}{}} & \rot{All} & \rot{Title} & \rot{Title} & \rot{All} & \rot{All} & \rot{Title} & \rot{Title} & \rot{Title} & \rot{Title} & \rot{All} & \rot{All} \\  \cline{5-15}
		\multicolumn{4}{c|}{} & \rot{\breakcell{Sliding\\window}} & \rot{\breakcell{Sliding\\window}} & \rot{\breakcell{Exponen-\\tial decay}} & \rot{\breakcell{Exponen-\\tial decay}} & \rot{\breakcell{Exponen-\\tial decay}} & \rot{\breakcell{Exponen-\\tial decay}} & \rot{\breakcell{Sliding\\window}} & \rot{\breakcell{Exponen-\\tial decay}} & \rot{\breakcell{Sliding\\window}} & \rot{\breakcell{Exponen-\\tial decay}} & \rot{\breakcell{Sliding\\window}} \\  \cline{5-15}
    \multicolumn{4}{c|}{} & \rot{HCF-IDF} & \rot{HCF-IDF} & \rot{HCF-IDF} & \rot{CF-IDF} & \rot{HCF-IDF} & \rot{CF-IDF} & \rot{CF-IDF} & \rot{LDA} & \rot{LDA} & \rot{LDA} & \rot{LDA} \\  \cline{5-15}
		\multicolumn{4}{c|}{} & 2. & 3. & 4. & 5. & 6. & 7. & 8. & 9. & 10. & 11. & 12. \\ \hhline{----*{10}{|=}|=|}
		1. & CF-IDF & Sliding window & All & .99 & .99 & .85 & .41 & .22 & \textbf{.00} & \textbf{.00} & \textbf{.00} & \textbf{.00} & \textbf{.00} & \textbf{.00} \\  \hline
    2. & HCF-IDF & Sliding window & All & \cellcolor[HTML]{9B9B9B} & .99 & .99 & .99 & .99 & \textbf{.00} & \textbf{.00} & \textbf{.00} & \textbf{.00} & \textbf{.00} & \textbf{.00} \\  \hline
		3. & HCF-IDF & Sliding window & Title & \cellcolor[HTML]{9B9B9B} & \cellcolor[HTML]{9B9B9B} & .99 & .99 & .99 & \textbf{.01} & \textbf{.00} & \textbf{.00} & \textbf{.00} & \textbf{.00} & \textbf{.00} \\  \hline
		4. & HCF-IDF & Exponential decay & Title & \cellcolor[HTML]{9B9B9B} & \cellcolor[HTML]{9B9B9B} & \cellcolor[HTML]{9B9B9B} & .99 & .99 & \textbf{.01} & \textbf{.00} & \textbf{.00} & \textbf{.00} & \textbf{.00} & \textbf{.00} \\  \hline
		5. & CF-IDF & Exponential decay & All & \cellcolor[HTML]{9B9B9B} & \cellcolor[HTML]{9B9B9B} & \cellcolor[HTML]{9B9B9B} & \cellcolor[HTML]{9B9B9B} & .99 & .05 & \textbf{.01} & \textbf{.00} & \textbf{.00} & \textbf{.00} & \textbf{.00} \\  \hline
		6. & HCF-IDF & Exponential decay & All & \cellcolor[HTML]{9B9B9B} & \cellcolor[HTML]{9B9B9B} & \cellcolor[HTML]{9B9B9B} & \cellcolor[HTML]{9B9B9B} & \cellcolor[HTML]{9B9B9B} & .17 & \textbf{.03} & \textbf{.00} & \textbf{.00} & \textbf{.00} & \textbf{.00} \\  \hline
		7. & CF-IDF & Exponential decay & Title & \cellcolor[HTML]{9B9B9B} & \cellcolor[HTML]{9B9B9B} & \cellcolor[HTML]{9B9B9B} & \cellcolor[HTML]{9B9B9B}& \cellcolor[HTML]{9B9B9B} & \cellcolor[HTML]{9B9B9B} & .99 & .85 & .18 & .34 & \textbf{.01} \\  \hline
		8. & CF-IDF & Sliding window & Title & \cellcolor[HTML]{9B9B9B} & \cellcolor[HTML]{9B9B9B} & \cellcolor[HTML]{9B9B9B} & \cellcolor[HTML]{9B9B9B} & \cellcolor[HTML]{9B9B9B} & \cellcolor[HTML]{9B9B9B} & \cellcolor[HTML]{9B9B9B} & .99 & .41 & .77 & \textbf{.05} \\  \hline
    9. & LDA & Exponential decay & Title & \cellcolor[HTML]{9B9B9B} & \cellcolor[HTML]{9B9B9B} & \cellcolor[HTML]{9B9B9B} & \cellcolor[HTML]{9B9B9B} & \cellcolor[HTML]{9B9B9B} & \cellcolor[HTML]{9B9B9B} & \cellcolor[HTML]{9B9B9B} & \cellcolor[HTML]{9B9B9B} & .99 & .99 & .99 \\  \hline
		10. & LDA & Sliding window & Title & \cellcolor[HTML]{9B9B9B} & \cellcolor[HTML]{9B9B9B} & \cellcolor[HTML]{9B9B9B} & \cellcolor[HTML]{9B9B9B} & \cellcolor[HTML]{9B9B9B} & \cellcolor[HTML]{9B9B9B} & \cellcolor[HTML]{9B9B9B} & \cellcolor[HTML]{9B9B9B} & \cellcolor[HTML]{9B9B9B} & .99 & .99 \\  \hline
		11. & LDA & Exponential decay & All & \cellcolor[HTML]{9B9B9B} & \cellcolor[HTML]{9B9B9B} & \cellcolor[HTML]{9B9B9B} & \cellcolor[HTML]{9B9B9B} & \cellcolor[HTML]{9B9B9B} & \cellcolor[HTML]{9B9B9B} & \cellcolor[HTML]{9B9B9B} & \cellcolor[HTML]{9B9B9B} & \cellcolor[HTML]{9B9B9B} & \cellcolor[HTML]{9B9B9B} & .98 \\  \hline
\end{tabular}
\label{tb:pairwise-ndcg}
\end{table*}
\subsection{Difference in experiment factors}
Subsequently, we analyze the results of nDCGs with respect to each factor. 
First, we apply Mendoza's test~\cite{mendoza:1980} which shows violations of sphericity against the factors \textit{Profiling Method} ($\chi^2(2) = 11.29$, $p = .00$), \textit{Profiling Method} $\times$ \textit{Decay Function} ($\chi^2(2) = 18.90$, $p = .00$), and \textit{Profiling Method} $\times$ \textit{Document Content} ($\chi^2(2) = 8.61$, $p = .01$). 
Thus, we run three-way repeated-measure ANOVA with a Greenhouse-Geisser correction of $\epsilon = .92$ for the factor \textit{Profiling Method}, $\epsilon = .87$ for the factor \textit{Profiling Method} $\times$ \textit{Decay Function}, and $\epsilon = .94$ for the factor \textit{Profiling Method} $\times$ \textit{Document Content}. 
Table~\ref{tb:three-way-anova-ndcg} shows the results of applying an ANOVA. 
$\eta^2$ indicates the effect size of each factor. 
For all the factors that make significant difference, we conduct a post-hoc analysis using Shaffer's MSRB Procedure. 
\begin{table}[!ht]
\centering
\small
\caption{Three-way repeated-measure ANOVA with Greenhouse-Geisser correction with F-ratio, $\eta^2$, and p-value for nDCG.}
\begin{tabular}{|l|r|r|r|} \hline
    \multicolumn{1}{|c|}{\textbf{Factor}} & \multicolumn{1}{c|}{\textbf{F}} & \multicolumn{1}{c|}{\textbf{$\eta^2$}} & \multicolumn{1}{c|}{\textbf{p}} \\ \hline\hline
		\textit{Profiling Method} & 58.42 & . & \textbf{.00} \\ \hline
    \textit{Decay Function} & 0.80 & . & .37 \\ \hline
    \textit{Document Content} & 6.33 & . & \textbf{.01} \\ \hline
    \textit{Profiling Method} $\times$ \textit{Decay Function} & 3.81 & . & \textbf{.03} \\ \hline
    \textit{Profiling Method} $\times$ \textit{Document Content} & 14.54 & . & \textbf{.00} \\ \hline
    \textit{Decay Function} $\times$ \textit{Document Content} & 3.57 & . & .06 \\ \hline
		\breakcell{\textit{Profiling Method} $\times$ \textit{Decay Function} $\times$ \\ \textit{Document Content}} & 3.09 & . & \textbf{.05} \\  \hline
\end{tabular}
\label{tb:three-way-anova-ndcg}
\end{table}
\paragraph*{\textbf{The factor \textit{Profiling Method}}}
Tables~\ref{tb:post-i-ndcg}(a), (b) and (c) show the MAPs, the post-hoc analysis for the factor \textit{Profiling Method}, and the effect size, respectively. 
Table~\ref{tb:post-i-ndcg}(a) presents the means and standard deviations of the three profiling methods. 
Table~\ref{tb:post-i-ndcg}(b) shows p-values of each pair. 
Since Table~\ref{tb:three-way-anova-ndcg} shows that the factor \textit{Profiling Method} has the largest effect size, we further compute the effect size using Cohen's d for each pair shown in Table~\ref{tb:post-i-ndcg}(c). 
The result shows that CF-IDF and HCF-IDF are superior to LDA. 
In contrast, there is no significant difference between CF-IDF and HCF-IDF, although MAP of HCF-IDF is slightly better than CF-IDF.
\begin{table}[!ht]
\centering
\small
\caption{nDCGs, Post-hoc analysis for the factor \textit{Profiling Method} using Shaffer's MSRB procedure, and effect size.}
\begin{tabular}{|c|c|c|}
		\multicolumn{3}{c}{\textbf{a) nDCGs}} \\  \hline
    \multicolumn{1}{|c|}{\textbf{Choice}} & \multicolumn{1}{c|}{\textbf{M}} & \multicolumn{1}{c|}{\textbf{SD}} \\  \hline\hline
    HCF-IDF & .53 & .32 \\  \hline
		CF-IDF & .48 & .32 \\  \hline
	  LDA & .32 & .32 \\  \hline
		\multicolumn{3}{c}{\textbf{b) Post-hoc analysis p-values}} \\  \hline
     & HCF-IDF & LDA \\  \hline \hline
	  CF-IDF & \textbf{.00} & \textbf{.00}  \\  \hline
    HCF-IDF & \cellcolor[HTML]{9B9B9B} & \textbf{.00} \\  \hline
		\multicolumn{3}{c}{\textbf{c) Effect size using Cohen's d}} \\  \hline
     & HCF-IDF & LDA \\  \hline \hline
	  CF-IDF & . & . \\  \hline
    HCF-IDF & \cellcolor[HTML]{9B9B9B} & . \\  \hline
\end{tabular}
\label{tb:post-i-ndcg}
\end{table}
\paragraph*{\textbf{The factor \textit{Document Content}}}
Table~\ref{tb:post-iii-ndcg} shows the post-hoc analysis for the factor \textit{Document Content}. 
It indicates that the recommender system works better when All (i\,.e., full-texts and titles) is taken into consideration for computing recommendations. 
\begin{table}[!ht]
\centering
\small
\caption{nDCGs and Post-hoc analysis for the factor \textit{Document Content} using Shaffer's MSRB procedure.}
\begin{tabular}{|c|c|c|}
		\multicolumn{3}{c}{\textbf{a) nDCGs}} \\  \hline
    \multicolumn{1}{|c|}{\textbf{Choice}} & \multicolumn{1}{c|}{\textbf{M}} & \multicolumn{1}{c|}{\textbf{SD}} \\  \hline\hline
		All & .46 & .34 \\  \hline
    Title & .42 & .31 \\  \hline
		\multicolumn{3}{c}{\textbf{b) Post-hoc analysis p-values}} \\  \hline
     & \multicolumn{2}{c|}{Title} \\  \hline \hline
	  All & \multicolumn{2}{c|}{\textbf{.01}}  \\  \hline
\end{tabular}
\label{tb:post-iii-ndcg}
\end{table}
\paragraph*{\textbf{The factor \textit{Profiling Method} $\times$ \textit{Decay Function}}}
Table~\ref{tb:anova-i-ii-ndcg} shows the results of ANOVA regarding the factor \textit{Profiling Method} when a choice of the factor \textit{Decay Function} is fixed and vice versa. 
Mendoza's test found a violation of sphericity in the factor \textit{Profiling Method} when Sliding window is used ($\chi^2(2) = 7.55$, $p=.02$) and Exponential decay is used ($\chi^2(2) = 10.74$, $p=.00$). 
Thus, we run a one-way repeated-measure ANOVA with Greenhouse-Geisser correction of $\eta = .94$ for the first row in Table~\ref{tb:anova-i-iii-ndcg} and $\eta = .92$ for the second row in Table~\ref{tb:anova-i-iii-ndcg}. 
We observe significant differences when a choice of the factor \textit{Decay Function} is fixed and when LDA is employed. 
The post-hoc analyses of them are shown in Table~\ref{tb:post-i-sw-ndcg}, Table~\ref{tb:post-i-e-ndcg}, and Table~\ref{tb:post-ii-t-ndcg}, respectively. 
In Table~\ref{tb:post-i-sw-ndcg} and Table~\ref{tb:post-i-e-ndcg}, a choice of the factor \textit{Decay Function} is fixed. 
Table~\ref{tb:post-ii-t-ndcg} shows the post-hoc analysis of the factor \textit{Decay Function} when LDA is employed. 
It indicates Exponential decay performs better than Sliding window for LDA. 
\begin{table}[!ht]
\centering
\small
\caption{ANOVA for \textit{Profiling Method} $\times$ \textit{Decay Function} interaction}
\begin{tabular}{|l|r|r|r|} \hline
    \multicolumn{1}{|c|}{\textbf{Factor}} & \multicolumn{1}{c|}{\textbf{F}} & \multicolumn{1}{c|}{\textbf{$\eta^2$}} & \multicolumn{1}{c|}{\textbf{p}} \\  \hline\hline
		\textit{Profiling Method} at Sliding window & 50.59 & . & \textbf{.00} \\  \hline
    \textit{Profiling Method} at Exponential decay & 27.92 & . & \textbf{.00} \\  \hline
    \textit{Decay Function} at CF-IDF & 2.79 & . & .10 \\  \hline
    \textit{Decay Function} at HCF-IDF & 1.78 & . & .18 \\  \hline
    \textit{Decay Function} at LDA & 4.90 & . & \textbf{.03} \\  \hline
\end{tabular}
\label{tb:anova-i-ii-ndcg}
\end{table}
\begin{table}[!ht]
\small
\centering
\caption{nDCGs and Post-hoc analysis for the factor \textit{Profiling Method} at Sliding window using Shaffer's MSRB procedure.}
\begin{tabular}{|c|c|c|}
    \multicolumn{3}{c}{\textbf{a) nDCGs}} \\  \hline
    \multicolumn{1}{|c|}{\textbf{Choice}} & \multicolumn{1}{c|}{\textbf{M}} & \multicolumn{1}{c|}{\textbf{SD}} \\  \hline\hline
    HCF-IDF & .55 & .33 \\  \hline
		CF-IDF & .50 & .32 \\  \hline
	  LDA & .30 & .32 \\  \hline
    \multicolumn{3}{c}{\textbf{b) Post-hoc analysis p-values}} \\  \hline
    & HCF-IDF & LDA \\  \hline \hline
	  CF-IDF & \textbf{.02} & \textbf{.00}  \\  \hline
    HCF-IDF & \cellcolor[HTML]{9B9B9B} & \textbf{.00} \\  \hline
\end{tabular}
\label{tb:post-i-sw-ndcg}
\end{table}
\begin{table}[!ht]
\small
\centering
\caption{nDCGs and Post-hoc analysis for the factor \textit{Profiling Method} at Exponential decay using Shaffer's MSRB procedure.}
\begin{tabular}{|c|c|c|}
		\multicolumn{3}{c}{\textbf{a) nDCGs}} \\  \hline
    \multicolumn{1}{|c|}{\textbf{Choice}} & \multicolumn{1}{c|}{\textbf{M}} & \multicolumn{1}{c|}{\textbf{SD}} \\  \hline\hline
    HCF-IDF & .51 & .30 \\  \hline
		CF-IDF & .46 & .31 \\  \hline
	  LDA & .34 & .31 \\  \hline
		\multicolumn{3}{c}{\textbf{b) Post-hoc analysis p-values}} \\  \hline
        & HCF-IDF & LDA \\  \hline \hline
	  CF-IDF & \textbf{.03} & \textbf{.00}  \\  \hline
    HCF-IDF & \cellcolor[HTML]{9B9B9B} & \textbf{.00} \\  \hline
\end{tabular}
\label{tb:post-i-e-ndcg}
\end{table}
\begin{table}[!ht]
\centering
\small
\caption{nDCGs and Post-hoc analysis for the factor \textit{Decay Function} at LDA using Shaffer's MSRB procedure.}
\begin{tabular}{|c|c|c|}
		\multicolumn{3}{c}{\textbf{a) nDCGs}} \\  \hline
    \multicolumn{1}{|c|}{\textbf{Choice}} & \multicolumn{1}{c|}{\textbf{M}} & \multicolumn{1}{c|}{\textbf{SD}} \\  \hline\hline
		Exponential decay & .34 & .31 \\  \hline
    Sliding window & .30 & .32 \\  \hline
		\multicolumn{3}{c}{\textbf{b) Post-hoc analysis p-value}} \\  \hline
        & \multicolumn{2}{c|}{Exponential decay} \\  \hline \hline
	  Sliding window & \multicolumn{2}{c|}{\textbf{.03}} \\  \hline
\end{tabular}
\label{tb:post-ii-t-ndcg}
\end{table}
\paragraph*{\textbf{The factor \textit{Profiling Method} $\times$ \textit{Document Content}}}
Table~\ref{tb:anova-i-iii-ndcg} shows the results of ANOVA regarding the factor \textit{Profiling Method} when a choice of the factor \textit{Document Content} is fixed and vice versa. 
We observe there are significant differences when a choice of the factor \textit{Document Content} is fixed and CF-IDF is employed. 
Mendoza's test found a violation of sphericity in the factor \textit{Profiling Method} when All is taken ($\chi^2(2) = 24.64$, $p=.00$). 
Thus, we run a one-way repeated-measure ANOVA with Greenhouse-Geisser correction of $\eta = .84$ for the second row in Table~\ref{tb:anova-i-iii-ndcg}. 
Table~\ref{tb:post-i-t-map} presents the post-hoc analysis when Title is employed. 
We see that HCF-IDF outperforms others with significant differences. 
Table~\ref{tb:post-i-f-map} shows the post-hoc analysis when All is chosen for the factor \textit{Document Content}. 
There is no significant difference between CF-IDF and HCF-IDF. 
Table~\ref{tb:post-iii-c-map} shows the post-hoc analysis of the factor \textit{Document Content} when CF-IDF is employed. 
It indicates that the strategies with CF-IDF and All significantly outperforms those with CF-IDF and Title. 
\begin{table}[!ht]
\centering
\small
\caption{ANOVA for \textit{Profiling Method} $\times$ \textit{Document Content} interaction}
\begin{tabular}{|l|r|r|r|} \hline
    \multicolumn{1}{|c|}{\textbf{Factor}} & \multicolumn{1}{c|}{\textbf{F}} & \multicolumn{1}{c|}{\textbf{$\eta^2$}} & \multicolumn{1}{c|}{\textbf{p}} \\  \hline\hline
		\textit{Profiling Method} at Title & 26.61 & . & \textbf{.00} \\  \hline
    \textit{Profiling Method} at All & 52.51 & . & \textbf{.00} \\  \hline
    \textit{Document Content} at CF-IDF & 30.81 & . & \textbf{.00} \\  \hline
    \textit{Document Content} at HCF-IDF & 0.31 & . & .58 \\  \hline
    \textit{Document Content} at LDA & 0.94 & . & .33 \\  \hline
\end{tabular}
\label{tb:anova-i-iii-ndcg}
\end{table}
\begin{table}[!ht]
\centering
\small
\caption{nDCGs and Post-hoc analysis for the factor \textit{Profiling Method} at Title using Shaffer's MSRB procedure.}
\begin{tabular}{|c|c|c|}
		\multicolumn{3}{c}{\textbf{a) nDCGs}} \\  \hline
    \multicolumn{1}{|c|}{\textbf{Choice}} & \multicolumn{1}{c|}{\textbf{M}} & \multicolumn{1}{c|}{\textbf{SD}} \\  \hline\hline
    HCF-IDF & .53 & .32 \\  \hline
		CF-IDF & .41 & .29 \\  \hline
	  LDA & .33 & .31 \\  \hline
		\multicolumn{3}{c}{\textbf{b) Post-hoc analysis p-values}} \\  \hline
        & HCF-IDF & LDA \\  \hline \hline
	  CF-IDF & \textbf{.00} & \textbf{.01}  \\  \hline
    HCF-IDF & \cellcolor[HTML]{9B9B9B} & \textbf{.00} \\  \hline
\end{tabular}
\label{tb:post-i-t-ndcg}
\end{table}
\begin{table}[!ht]
\centering
\small
\caption{nDCG and Post-hoc analysis for the factor \textit{Profiling Method} at All using Shaffer's MSRB procedure.}
\begin{tabular}{|c|c|c|}
		\multicolumn{3}{c}{\textbf{a) nDCGs}} \\  \hline
    \multicolumn{1}{|c|}{\textbf{Choice}} & \multicolumn{1}{c|}{\textbf{M}} & \multicolumn{1}{c|}{\textbf{SD}} \\  \hline\hline
		CF-IDF & .56 & .33\\  \hline
    HCF-IDF & .53 & .34 \\  \hline
	  LDA & .30 & .32 \\  \hline
		\multicolumn{3}{c}{\textbf{b) Post-hoc analysis p-values}} \\  \hline
        & HCF-IDF & LDA \\  \hline \hline
	  CF-IDF & .18 & \textbf{.00}  \\  \hline
    HCF-IDF & \cellcolor[HTML]{9B9B9B} & \textbf{.00} \\  \hline
\end{tabular}
\label{tb:post-i-f-ndcg}
\end{table}
\paragraph*{\textbf{The factor \textit{Decay Function} $\times$ \textit{Document Content}}}
Table~\ref{tb:anova-ii-iii-ndcg} shows the results of ANOVA regarding the factor \textit{Decay Function} when a choice of the factor \textit{Document Content} is fixed and vice versa. 
According to Table~\ref{tb:anova-ii-iii-ndcg}, there is a significant difference among the factor \textit{Document Content}, when Sliding window is used. 
The nDCGs and post-hoc analysis of it are shown in Tables~\ref{tb:post-iii-s-ndcg}(a) and (b). 
It indicates that All significantly enhances the performance of the recommender system when Sliding window is used. 
\begin{table}[!ht]
\centering
\small
\caption{ANOVA for \textit{Decay Function} $\times$ \textit{Document Content} interaction}
\begin{tabular}{|l|r|r|r|} \hline
    \multicolumn{1}{|c|}{\textbf{Factor}} & \multicolumn{1}{c|}{\textbf{F}} & \multicolumn{1}{c|}{\textbf{$\eta^2$}} & \multicolumn{1}{c|}{\textbf{p}} \\  \hline\hline
		{\textit{Decay Function}} at Title & 0.06 & . & .81 \\  \hline
    {\textit{Decay Function}} at All & 2.28 & . & .13 \\  \hline
    \textit{Document Content} at Sliding window & 9.96 & . & \textbf{.00} \\  \hline
    \textit{Document Content} at Exponential decay & 1.19 & . & .28 \\  \hline
\end{tabular}
\label{tb:anova-ii-iii-ndcg}
\end{table}
\begin{table}[!ht]
\centering
\small
\caption{nDCGs and Post-hoc analysis for the factor \textit{Document Content} at Sliding window using Shaffer's MSRB procedure.}
\begin{tabular}{|c|c|c|}
		\multicolumn{3}{c}{\textbf{a) nDCGs}} \\  \hline
    \multicolumn{1}{|c|}{\textbf{Choice}} & \multicolumn{1}{c|}{\textbf{M}} & \multicolumn{1}{c|}{\textbf{SD}} \\  \hline\hline
		All & .48 & .36 \\  \hline
    Title & .42 & .32 \\  \hline
		\multicolumn{3}{c}{\textbf{b) Post-hoc analysis p-value}} \\  \hline
        & \multicolumn{2}{c|}{All} \\  \hline \hline
	  Title & \multicolumn{2}{c|}{\textbf{.00}} \\  \hline
\end{tabular}
\label{tb:post-iii-s-ndcg}
\end{table}
\section{Correlation with the number of tweets, the number of concepts, the number of concepts per tweet, and the percentage of tweets containing at least one concept}
We examine whether the recommendation performance of the twelve strategies measured by the rankscore have correlations with the number of tweets, the number of concepts, the number of concepts per tweet, and the percentage of tweets containing at least one concept. 
We compute Pearson product-moment correlation coefficient and Kendall rank correlation coefficient between the rankscores of each of the twelve strategies and each of the number of tweets, the number of concepts, the number of concepts per tweet, and the percentage of tweets containing at least one concept. 
Table~\ref{tb:correlation} provides the results. 
\begin{table*}[!ht]
\small
\caption{Correlation coefficients of each of twelve strategies with the number of tweets, the number of concepts, the number of concepts per tweet, and the percentage of tweets containing at least one concept. We calculate correlation coefficients using Pearson product-moment correlation coefficient and Kendall rank correlation coefficient. Strategies are sorted by rankscores as shown in Table~\ref{tb:single-all}. In parentheses, p-values are given and marked in bold font, if < .05. Strategies are sorted by rankscores as shown in Table~\ref{tb:single-all}.}
\begin{tabular}{|c|l|l|l||r|r||r|r||r|r|} \hline
		\multirow{2}{*}{} & \multicolumn{3}{c||}{\textbf{Strategy}} & \multicolumn{2}{c||}{\textbf{\# of tweets}} & \multicolumn{2}{c||}{\textbf{\# of concepts}} & \multicolumn{2}{c|}{\textbf{\# of concepts per tweet}} \\  \cline{2-10}
		& \multicolumn{1}{c|}{\breakcell{\textbf{Profiling}\\ \textbf{Method}}} & \multicolumn{1}{c|}{\breakcell{\textbf{Decay}\\ \textbf{Function}}} & \multicolumn{1}{c||}{\breakcell{\textbf{Document} \\ \textbf{Content}}} & \multicolumn{1}{c|}{\textbf{Pearson}} & \multicolumn{1}{c||}{\textbf{Kendall}} & \multicolumn{1}{c|}{\textbf{Pearson}} & \multicolumn{1}{c||}{\textbf{Kendall}} & \multicolumn{1}{c|}{\textbf{Pearson}} & \multicolumn{1}{c|}{\textbf{Kendall}} \\  \hline\hline	
		1. & CF-IDF & Sliding window & All & -.02 (.82) & -.01 (.84) & .00 (.98) & -.01 (.91) & -.06 (.54) & -.01 (.82) \\  \hline
		2. & HCF-IDF & Sliding window & All & -.02 (.85) & -.02 (.77) & .00 (.99) & .00 (.96) & .08 (.38) & .05 (.46) \\  \hline
		3. & HCF-IDF & Sliding window & Title & -.12 (.20) & -.07 (.26) & -.07 (.41) & -.04 (.51) & .04 (.67) & .05 (.38) \\  \hline
		4. & HCF-IDF & Exponential decay & Title & -.01 (.94) & -.03 (.68) & -.03 (.74) & -.04 (.57) & -.10 (.26) & -.09 (.15) \\  \hline
		5. & CF-IDF & Exponential decay & All & .02 (.83) & .02 (.80) & .00 (.93) & .01 (.92) & -.09 (.35) & -.08 (.23) \\  \hline
		6. & HCF-IDF & Exponential decay & All & .03 (.72) & .02 (.71) & .04 (.63) & .03 (.68) & -.06 (.49) & -.06 (.32) \\  \hline
		7. & CF-IDF & Exponential decay & Title & .17 (.07) & .08 (.19) & .13 (.14) & .07 (.25) & -.13 (.17) & -.05 (40) \\  \hline
		8. & CF-IDF & Sliding window & Title & .14 (.12) & .09 (.17) & .13 (.15) & .06 (.32) & -.10 (.26) & -.03 (.65) \\  \hline
		9. & LDA & Exponential decay & Title & .15 (.09) & .12 (\textbf{.05}) & .14 (.12) & .11 (.07) & .02 (.78) & .00 (.99) \\  \hline
		10. & LDA & Sliding window & Title & .18 (.05) & .08 (.24) & .14 (.12) & .06 (.34) & -.03 (.73) & -.01 (.89) \\  \hline
		11. & LDA & Exponential decay & All & -.10 (.28) & -.13 (\textbf{.03}) & -.07 (.46) & -.12 (\textbf{.05}) & .06 (.48) & .03 (.65) \\  \hline
		12. &	LDA & Sliding window & All & -.06 (.50) & -.08 (.25) & -.04 (.66) & -.07 (.28) & -.03 (.76) & .01 (.93) \\  \hline
\end{tabular}
\begin{tabular}{|c|l|l|l||r|r|} \hline
		\multirow{2}{*}{} & \multicolumn{3}{c||}{\textbf{Strategy}} & \multicolumn{2}{c|}{\textbf{percentage of tweets with concepts}} \\  \cline{2-6}
		& \multicolumn{1}{c|}{\breakcell{\textbf{Profiling}\\ \textbf{Method}}} & \multicolumn{1}{c|}{\breakcell{\textbf{Decay}\\ \textbf{Function}}} & \multicolumn{1}{c||}{\breakcell{\textbf{Document} \\ \textbf{Content}}} & \multicolumn{1}{c|}{\textbf{Pearson}} & \multicolumn{1}{c|}{\textbf{Kendall}} \\  \hline\hline	
		1. & CF-IDF & Sliding window & All & -.02 (.79) & .01 (.92) \\  \hline
		2. & HCF-IDF & Sliding window & All & .11 (.22) & .06 (.32) \\  \hline
		3. & HCF-IDF & Sliding window & Title & .06 (.49) & .06 (.31) \\  \hline
		4. & HCF-IDF & Exponential decay & Title & -.11 (.24) & -.10 (.11) \\  \hline
		5. & CF-IDF & Exponential decay & All & -.08 (.38) & -.07 (.24) \\  \hline
		6. & HCF-IDF & Exponential decay & All & -.05 (.62) & -.07 (.24) \\  \hline
		7. & CF-IDF & Exponential decay & Title & -.14 (.12) & -.08 (.20) \\  \hline
		8. & CF-IDF & Sliding window & Title & -.13 (.16) & -.04 (.50) \\  \hline
		9. & LDA & Exponential decay & Title & .03 (.76) & .00 (.97) \\  \hline
		10. & LDA & Sliding window & Title & -.03 (.72) & -.01 (.82) \\  \hline
		11. & LDA & Exponential decay & All & .07 (.42) & .04 (.54) \\  \hline
		12. &	LDA & Sliding window & All & -.03 (.71) & -.01 (.92) \\  \hline
\end{tabular}
\label{tb:correlation}
\end{table*}
\section{User click rates}
In this section, we describe the details of the analysis regarding click rates. 
\subsection{Three-way repeated ANOVA for click rates}
Table~\ref{tb:click-rate-strategy} shows average click rates of each strategy. 
Click rates differ depending on each strategy. 
In order to reveal effects on click rates from factors, we run three-way repeated ANOVA. 
First, we apply Mendoza's test~\cite{mendoza:1980} which showed violations of sphericity against the factors \textit{Profiling Method} ($\chi^2(2) = 12.24$, $p = .00$), \textit{Profiling Method} $\times$ \textit{Decay Function} ($\chi^2(2) = 8.44$, $p = .01$), and \textit{Profiling Method} $\times$ \textit{Document Content} ($\chi^2(2) = 13.57$, $p = .00$). 
Thus, we run three-way repeated-measure ANOVA with a Greenhouse-Geisser correction of $\epsilon = .91$ for the factor \textit{Profiling Method}, $\epsilon = .94$ for the factor \textit{Profiling Method} $\times$ \textit{Decay Function}, and $\epsilon = .90$ for the factor \textit{Profiling Method} $\times$ \textit{Document Content}. 
Table~\ref{tb:three-way-anova-click} shows the results of an ANOVA with F-ratio, $\eta^2$ and p-value. 
$\eta^2$ indicates the effect size of each factor. 
The effect size is small when $\eta^2 > .02$, medium when $\eta^2 > .13$, and large when $\eta^2 > .26$. 
\begin{table}[!ht]
\small
\centering
\caption{Average click rates on the PDF files. In parentheses, the standard deviations are shown. Strategies are sorted by rankscores as shown in Table~\ref{tb:single-all}.}
\begin{tabular}{|c|l|l|l||r|} \hline
		\multirow{2}{*}{} & \multicolumn{3}{c||}{\textbf{Strategy}} & \multicolumn{1}{c|}{\textbf{Click rate}} \\  \cline{2-5}
		& \multicolumn{1}{c|}{\breakcell{\textbf{Profiling}\\ \textbf{Method}}} & \multicolumn{1}{c|}{\breakcell{\textbf{Decay}\\ \textbf{Function}}} & \multicolumn{1}{c||}{\breakcell{\textbf{Con-} \\ \textbf{tent}}} & \multicolumn{1}{c|}{\textbf{Rate}} \\  \hline\hline
		1. & CF-IDF & Sliding window & All & 10.73\% (24.73) \\  \hline
		2. & HCF-IDF & Sliding window & All & 10.08\% (23.94) \\  \hline
		3. & HCF-IDF & Sliding window & Title & 9.11\% (23.22) \\  \hline
		4. & HCF-IDF & Exponential decay & Title & 7.64\% (17.28) \\  \hline
		5. & CF-IDF & Exponential decay & All & 9.11\% (22.21) \\  \hline
		6. & HCF-IDF & Exponential decay & All & 8.29\% (20.31) \\  \hline
		7. & CF-IDF & Exponential decay & Title & 8.94\% (20.03) \\  \hline
		8. & CF-IDF & Sliding window & Title & 9.59\% (22.81) \\  \hline
		9. & LDA & Exponential decay & Title & 4.23\% (13.12) \\  \hline
		10. & LDA & Sliding window & Title & 4.72\% (15.38) \\  \hline
		11. & LDA & Exponential decay & All & 9.27\% (21.47) \\  \hline
		12. & LDA & Sliding window & All & 5.37\% (16.41) \\  \hline
\end{tabular}
\label{tb:click-rate-strategy}
\end{table}
\begin{table}[!ht]
\centering
\small
\caption{Three-way repeated-measure ANOVA for click rates with Greenhouse-Geisser correction with F-ratio, $\eta^2$ and p-value.}
\begin{tabular}{|l|r|r|r|} \hline
    \multicolumn{1}{|c|}{\textbf{Factor}} & \multicolumn{1}{c|}{\textbf{F}} & \multicolumn{1}{c|}{\textbf{$\eta^2$}} & \multicolumn{1}{c|}{\textbf{p}} \\  \hline\hline
		\textit{Profiling Method} & 5.23 & .04 & \textbf{.01} \\  \hline
    \textit{Decay Function} & 0.23 & .00 & .64 \\  \hline
    \textit{Document Content} & 2.60 & .02 & .11 \\  \hline
    \textit{Profiling Method} $\times$ \textit{Decay Function} & 2.71 & .02 & .07 \\  \hline
    \textit{Profiling Method} $\times$ \textit{Document Content} & 1.16 & .01 & .32 \\  \hline
    \textit{Decay Function} $\times$ \textit{Document Content} & 1.00 & .01 & .10 \\  \hline
		\breakcell{\textit{Profiling Method} $\times$ \textit{Decay Function} $\times$ \\ \textit{Document Content}} & 2.31 & .02 & \textbf{.04} \\  \hline
\end{tabular}
\label{tb:three-way-anova-click}
\end{table}

According to~\ref{tb:three-way-anova-click}, the factor \textit{Profiling Method} makes significant difference on click rates. 
We applied again a post-hoc using Shaffer's MSRB Procedure for the factor \textit{Profiling Method}. 
Table~\ref{tb:post-i-click} provides the result of the post-hoc analysis for the factor \textit{Profiling Method}. 
It indicates that click rates of CF-IDF and HCF-IDF are significantly higher than those of LDA. 
\begin{table}[!ht]
\small
\centering
\caption{Click rates, Post-hoc analysis for the factor \textit{Profiling Method} using Shaffer's MSRB procedure, and effect size.}
\begin{tabular}{|c|c|c|}
		\multicolumn{3}{c}{\textbf{a) Click rates}} \\  \hline
    \multicolumn{1}{|c|}{\textbf{Choice}} & \multicolumn{1}{c|}{\textbf{M}} & \multicolumn{1}{c|}{\textbf{SD}} \\  \hline\hline
    HCF-IDF & .09 & .21 \\  \hline
		CF-IDF & .10 & .22 \\  \hline
	  LDA & .06 & .17 \\  \hline
		\multicolumn{3}{c}{\textbf{b) Post-hoc analysis p-values}} \\  \hline
     & HCF-IDF & LDA \\  \hline \hline
	  CF-IDF & .42 & \textbf{.01}  \\  \hline
    HCF-IDF & \cellcolor[HTML]{9B9B9B} & \textbf{.03} \\  \hline
		\multicolumn{3}{c}{\textbf{c) Effect size using Cohen's d}} \\  \hline
     & HCF-IDF & LDA \\  \hline \hline
	  CF-IDF & .05 & .20 \\  \hline
    HCF-IDF & \cellcolor[HTML]{9B9B9B} & .16 \\  \hline
\end{tabular}
\label{tb:post-i-click}
\end{table}
\begin{table*}[!ht]
\centering
\small
\caption{Correlation coefficients of each of twelve strategies between click rates and rankscores. We calculate correlation coefficients using Pearson product-moment correlation coefficient and Kendall rank correlation coefficient. Strategies are sorted by rankscores as shown in Table~\ref{tb:single-all}. In parentheses, p-values are given and marked in bold font, if < .05. Strategies are sorted by rankscores as shown in Table~\ref{tb:single-all}.}
\begin{tabular}{|c|l|l|l||r|r|} \hline
		\multirow{2}{*}{} & \multicolumn{3}{c||}{\textbf{Strategy}} & \multicolumn{2}{c|}{\textbf{Correlation with click rates}} \\  \cline{2-6}
		 & \multicolumn{1}{c|}{\breakcell{\textbf{Profiling}\\ \textbf{Method}}} & \multicolumn{1}{c|}{\breakcell{\textbf{Decay}\\ \textbf{Function}}} & \multicolumn{1}{c||}{\breakcell{\textbf{Con-} \\ \textbf{tent}}} & \multicolumn{1}{c|}{\textbf{Pearson}} & \multicolumn{1}{c|}{\textbf{Kendall}} \\  \hline\hline
		1. & CF-IDF & Sliding window & All & .02 (.82) & -.03 (.68) \\  \hline
		2. & HCF-IDF & Sliding window & All & .07 (.42) & .05 (.47) \\  \hline
		3. & HCF-IDF & Sliding window & Title & -.06 (.52) & -.07 (.34) \\  \hline
		4. & HCF-IDF & Exponential decay & Title & .05 (.61) & .05 (.51) \\  \hline
		5. & CF-IDF & Exponential decay & All & .12 (.19) & .05 (.51) \\  \hline
		6. & HCF-IDF & Exponential decay & All & -.02 (.86) & -.02 (.78) \\  \hline
		7. & CF-IDF & Exponential decay & Title & .04 (.67) & .01 (.85) \\  \hline
		8. & CF-IDF & Sliding window & Title & .05 (.57) & .00 (.96) \\  \hline
		9. & LDA & Exponential decay & Title & .02 (.82) & -.02 (.80) \\  \hline
		10. & LDA & Sliding window & Title & -.01 (.90) & .02 (.79) \\  \hline
		11. & LDA & Exponential decay & All & .13 (.16) & .18 (\textbf{.02}) \\ \hline
		12. & LDA & Sliding window & All & .22 (\textbf{.01}) & .17 (\textbf{.03}) \\  \hline
\end{tabular}
\label{tb:correlation-click-rankscore}
\end{table*}
\subsection{Correlation with rankscores}
In addition, we investigate the correlation between click rates and rankscores with respect to each strategy. 
We use Pearson product-moment correlation coefficient as well as Kendall rank correlation coefficient. 
Table~\ref{tb:correlation-click-rankscore} reports the result of the analysis. 
\subsection{Precision for clicked PDF files}
Table~\ref{tb:click-rate-strategy-interesting} shows the average precision for clicked PDF files. 
It is equal to the probability that a participant evaluate a recommended item ``interesting" when clicking it. 
We observe that the precisions for the strategies HCF-IDF it are high even if recommendations are made with only titles. 
\begin{table}[!ht]
\small
\centering
\caption{Precision for clicked PDF files.}
\begin{tabular}{|c|l|l|l||r|} \hline
		\multirow{2}{*}{} & \multicolumn{3}{c||}{\textbf{Strategy}} & \multicolumn{1}{c|}{\textbf{Click rate}} \\  \cline{2-5}
		& \multicolumn{1}{c|}{\breakcell{\textbf{Profiling}\\ \textbf{Method}}} & \multicolumn{1}{c|}{\breakcell{\textbf{Decay}\\ \textbf{Function}}} & \multicolumn{1}{c||}{\breakcell{\textbf{Con-} \\ \textbf{tent}}} & \multicolumn{1}{c|}{} \\  \hline\hline
		1. & CF-IDF & Sliding window & All & 71.21\% \\  \hline
		2. & HCF-IDF & Sliding window & All & 64.52\% \\  \hline
		3. & HCF-IDF & Sliding window & Title & 55.36\% \\  \hline
		4. & HCF-IDF & Exponential decay & Title & 68.09\% \\  \hline
		5. & CF-IDF & Exponential decay & All & 71.43\% \\  \hline
		6. & HCF-IDF & Exponential decay & All & 60.78\% \\  \hline
		7. & CF-IDF & Exponential decay & Title & 47.27\% \\  \hline
		8. & CF-IDF & Sliding window & Title & 49.15\% \\  \hline
		9. & LDA & Exponential decay & Title & 42.31\% \\  \hline
		10. & LDA & Sliding window & Title & 37.93\% \\  \hline
		11. & LDA & Exponential decay & All & 43.86\% \\  \hline
		12. & LDA & Sliding window & All & 48.48\% \\  \hline
\end{tabular}
\label{tb:click-rate-strategy-interesting}
\end{table}
\section{Demographic Factor}
\label{sec:app-demographic-factor}
For each demographic factor, we first apply Mendoza's test. 
Subsequently, we conduct a mixed ANOVA test with one between subject factor (i.\,e., demographic factor) and one within subject factor (i.\,e., strategy), adjusted by Green-house-Geisser's epsilon. 
In addition, we provide the post-hoc analyses. 
However, we omit the post-hoc analysis of the factor strategy for the sake of brevity, because it is not so different from the result of the one-way repeated-measure ANOVA shown in Table~\ref{tb:pairwise}. 
\paragraph*{\textbf{Gender}} 
Mendoza's test found a violation of sphericity in the factor strategy ($\chi^2(131) = 489.39$, $p = .00$) when comparing the male ($n = 96$) and female ($n = 27$) participants.
Table~\ref{tb:gender} shows the result of an ANOVA with a Greenhouse-Geisser correction of $\epsilon = .60$. 
According to Table~\ref{tb:gender}, we see a significant difference between males and females. 
The post-hoc analysis is shown in Table~\ref{tb:post-gender}. 
We observe female participants are more likely to evaluate recommended items as interesting than males. 
However, the factor gender does no make any difference about how each of the twelve strategies performs compared to the other strategies, because there is no significant difference in the factor gender $\times$ strategy. 
\begin{table}[!ht]
\centering
\small
\caption{Mixed ANOVA with a between subject factor \textit{Gender} and a within subject factor \textit{Strategy} Greenhouse-Geisser correction with F-ratio, effect size $\eta^2$, and p-value.}
\begin{tabular}{|l|r|r|r|} \hline
    \multicolumn{1}{|c|}{\textbf{Factor}} & \multicolumn{1}{c|}{\textbf{F}} & \multicolumn{1}{c|}{\textbf{$\eta^2$}} & \multicolumn{1}{c|}{\textbf{p}} \\  \hline\hline
	  Gender & 9.69 & .08 & \textbf{.00} \\  \hline
    Strategy & 16.58 & .14 & \textbf{.00} \\  \hline
    Gender $\times$ Strategy & 1.11 & .01 & .36 \\  \hline
\end{tabular}
\label{tb:gender}
\end{table}
\begin{table}[!ht]
\centering
\small
\caption{Rankscores and Post-hoc analysis for the factor \textit{Gender} using Shaffer's MSRB procedure.}
\begin{tabular}{|c|c|c|}
		\multicolumn{3}{c}{\textbf{a) Rankscores}} \\  \hline
    \multicolumn{1}{|c|}{\textbf{Degree}} & \multicolumn{1}{c|}{\textbf{M}} & \multicolumn{1}{c|}{\textbf{SD}} \\  \hline\hline
		\multicolumn{1}{|c|}{male} & .42 & .32 \\  \hline
    \multicolumn{1}{|c|}{female} & .54 & .35 \\  \hline
		\multicolumn{3}{c}{\textbf{b) Post-hoc analysis p-values}} \\  \hline
        & \multicolumn{2}{c|}{female} \\  \hline \hline
	  male & \multicolumn{2}{c|}{\textbf{.00}} \\  \hline
\end{tabular}
\label{tb:post-gender}
\end{table}
\paragraph*{\textbf{Age}}
On average, participants are $32.90$ years old (SD: $7.36$). 
We divide participants into three groups for an ANOVA (group 1: participants who are $> 29$ years old ($n = 42$), group 2: $<= 29$ and $> 38$ years old ($n = 49$), group 3: $<= 38$ years old ($n = 32$)). 
We set those thresholds to make three groups have the almost same number of participants. 
Mendoza's test found a violation of sphericity in the strategies ($\chi^2(197) = 504.35$, $p = .00$). 
Table~\ref{tb:age} shows the result of an ANOVA with a Greenhouse-Geisser correction of $\epsilon = .60$. 
It indicates that the age of participants has no effect on the performance of the different strategies. 
\begin{table}[!ht]
\centering
\small
\caption{Mixed ANOVA with a between subject factor \textit{Age} and a within subject factor \textit{Strategy} Greenhouse-Geisser correction with F-ratio, effect size $\eta^2$, and p-value.}
\begin{tabular}{|l|r|r|r|} \hline
    \multicolumn{1}{|c|}{\textbf{Factor}} & \multicolumn{1}{c|}{\textbf{F}} & \multicolumn{1}{c|}{\textbf{$\eta^2$}} & \multicolumn{1}{c|}{\textbf{p}} \\  \hline\hline
	  Age & 2.06 & .03 & .13 \\  \hline
    Strategy & 14.82 & .12 & \textbf{.00} \\  \hline
    Age $\times$ Strategy & 0.69 & .01 & .77 \\  \hline
\end{tabular}
\label{tb:age}
\end{table}
\paragraph*{\textbf{Highest Academic Degree}}
We have participants whose highest academic degree is Bachelor ($n = 21$), Master ($n = 58$), PhD ($n = 32$), and lecturer/professor ($n = 12$). 
Mendoza's test found a violation of sphericity in the strategies when comparing the distributions among the factors ($\chi^2(263) = 653.03$, $p = .00$). 
Table~\ref{tb:degree} shows the result of applying an ANOVA with a Greenhouse-Geisser correction of $\epsilon = .60$.
According to Table~\ref{tb:degree}, we see a significant difference among participants grouped by their highest academic degrees. 
The post-hoc analysis is shown in Table~\ref{tb:post-degree}. 
We observe that participants whose highest academic degree is Bachelor are more likely to evaluate recommended items as interesting than those whose highest academic degree is lecturer/professor. 
\begin{table}[!ht]
\centering
\small
\caption{Mixed ANOVA with a between subject factor \textit{Highest Academic Degree} and a within subject factor \textit{Strategy} Greenhouse-Geisser correction with F-ratio, effect size $\eta^2$, and p-value.}
\begin{tabular}{|l|r|r|r|} \hline
    \multicolumn{1}{|c|}{\textbf{Factor}} & \multicolumn{1}{c|}{\textbf{F}} & \multicolumn{1}{c|}{\textbf{$\eta^2$}} & \multicolumn{1}{c|}{\textbf{p}} \\  \hline\hline
	  Highest Academic Degree & 3.38 & .09 & \textbf{.02} \\  \hline
    Strategy & 16.02 & .13 & \textbf{.00} \\  \hline
    Highest Academic Degree $\times$ Strategy & 0.77 & .02 & .75 \\  \hline
\end{tabular}
\label{tb:degree}
\end{table}
\begin{table}[!ht]
\centering
\small
\caption{Rankscores and Post-hoc analysis for the factor \textit{Highest Academic Degree} using Shaffer's MSRB procedure.}
\begin{tabular}{|c|c|c|c|}
		\multicolumn{4}{c}{\textbf{a) Rankscores}} \\  \hline
    \multicolumn{2}{|c|}{\textbf{Degree}} & \multicolumn{1}{c|}{\textbf{M}} & \multicolumn{1}{c|}{\textbf{SD}} \\  \hline\hline
		\multicolumn{2}{|c|}{Bachelor} & .53 & .30 \\  \hline
    \multicolumn{2}{|c|}{Master} & .43 & .33 \\  \hline
		\multicolumn{2}{|c|}{PhD} & .44 & .33 \\  \hline
		\multicolumn{2}{|c|}{lecturer/professor} & .32 & .28 \\  \hline
		\multicolumn{4}{c}{\textbf{b) Post-hoc analysis p-values}} \\  \hline
        & Master & PhD & lecturer/professor \\  \hline \hline
	  Bachelor & .20 & .21 & \textbf{.01} \\  \hline
		Master & \cellcolor[HTML]{9B9B9B} & .72 & .21 \\  \hline
		PhD & \cellcolor[HTML]{9B9B9B} & \cellcolor[HTML]{9B9B9B} & .09 \\  \hline
\end{tabular}
\label{tb:post-degree}
\end{table}
\paragraph*{\textbf{Major}}
In the experiment, participants provide information about their majors. 
We manually classify participants into the two groups: participants whose major is economics ($n = 92$) and others ($n = 31$). 
Mendoza's test found a violation of sphericity in the strategies for these two groups ($\chi^2(131) = 466.90$, $p = .00$). 
Table~\ref{tb:major} shows the result of an ANOVA with a Greenhouse-Geisser correction of $\epsilon = .60$. 
It indicates that the major of participants has no effect on the performance of the different strategies. 
\begin{table}[!ht]
\centering
\small
\caption{Mixed ANOVA with a between subject factor \textit{Major} and a within subject factor \textit{Strategy} Greenhouse-Geisser correction with F-ratio, effect size $\eta^2$, and p-value.}
\begin{tabular}{|l|r|r|r|} \hline
    \multicolumn{1}{|c|}{\textbf{Factor}} & \multicolumn{1}{c|}{\textbf{F}} & \multicolumn{1}{c|}{\textbf{$\eta^2$}} & \multicolumn{1}{c|}{\textbf{p}} \\  \hline\hline
	  Major & 0.01 & .00 & .94 \\  \hline
    Strategy & 16.41 & .14 & \textbf{.00} \\  \hline
    Major $\times$ Strategy & 1.73 & .01 & .10 \\  \hline
\end{tabular}
\label{tb:major}
\end{table}
\paragraph*{\textbf{Years of Profession}}
On average, participants work in their fields for $7.85$ years (SD:~$6.85$). 
We divide participants into three groups for an ANOVA (group 1: participants who work for $> 5$ years ($n = 44$), group 2: $<= 5$ and $> 10$ years ($n = 34$), group 3: $<= 10$ years ($n = 44$)). 
We set those thresholds to make three groups have the almost same number of participants. 
Mendoza's test found a violation of sphericity in the strategies ($\chi^2(197) = 541.67$, $p = .00$). 
Table~\ref{tb:work} shows the result of an ANOVA with a Greenhouse-Geisser correction of $\epsilon = .60$. 
It indicates that how long participants have worked in their fields has no effect on the performance of the different strategies. 
\begin{table}[!ht]
\centering
\small
\caption{Mixed ANOVA with a between subject factor \textit{Years of Profession} and a within subject factor \textit{Strategy} Greenhouse-Geisser correction with F-ratio, effect size $\eta^2$, and p-value.}
\begin{tabular}{|l|r|r|r|} \hline
    \multicolumn{1}{|c|}{\textbf{Factor}} & \multicolumn{1}{c|}{\textbf{F}} & \multicolumn{1}{c|}{\textbf{$\eta^2$}} & \multicolumn{1}{c|}{\textbf{p}} \\  \hline\hline
	  Years of Profession & 0.13 & .00 & .88 \\  \hline
    Strategy & 21.70 & .18 & \textbf{.00} \\  \hline
    Years of Profession $\times$ Strategy & 0.80 & .01 & .66 \\  \hline
\end{tabular}
\label{tb:duration}
\end{table}
\paragraph*{\textbf{Employment Type}}
We have participants who work in academia ($n = 83$) and industry ($n = 40$). Mendoza's test found a violation of sphericity in the strategies ($\chi^2(131) = 472.14$, $p = .00$). Table~\ref{tb:work} shows the result of an ANOVA with a Greenhouse-Geisser correction of $\epsilon = .60$. 
It indicates that the employment type of participants has no effect on the performance of the different strategies. 
\begin{table}[!ht]
\centering
\small
\caption{Mixed ANOVA with a between subject factor \textit{Employment Type} and a within subject factor \textit{Strategy} Greenhouse-Geisser correction with F-ratio, effect size $\eta^2$, and p-value.}
\begin{tabular}{|l|r|r|r|} \hline
    \multicolumn{1}{|c|}{\textbf{Factor}} & \multicolumn{1}{c|}{\textbf{F}} & \multicolumn{1}{c|}{\textbf{$\eta^2$}} & \multicolumn{1}{c|}{\textbf{p}} \\  \hline\hline
	  Employment Type & 0.35 & .00 & .55 \\  \hline
    Strategy & 18.05 & .15 & \textbf{.00} \\  \hline
    Employment Type $\times$ Strategy & 0.97 & .01 & .45 \\  \hline
\end{tabular}
\label{tb:work}
\end{table}

\remove{
\section{ToDo}
\begin{itemize}
	\item DB Table userPublication: compare performance at each rank computed by algorithm and at each position
	\item (figure precision at each rank, x-axis: precision, y-axis: each rank(1 to 5))
	\item report Likert scale by mode
	\item correlation analysis for the demographic factor age and years of profession
\end{itemize}
}

\balancecolumns
\end{document}